\documentclass[aps,prb,twocolumn,superscriptaddress,amsmath,amssymb]{revtex4}

\usepackage{graphicx}

\begin{document}


\title{A new universal ratio in Random Matrix Theory
and chaotic to integrable transition
in Type-I and Type-II hybrid Sachdev-Ye-Kitaev models}
\author{Fadi Sun}
\affiliation{Institute for Theoretical Sciences,
Westlake University, Hangzhou, 310024, Zhejiang, China}
\affiliation{Tsung-Dao Lee Institute, Shanghai 200240, China}
\affiliation{Department of Physics and Astronomy, Mississippi State University, MS 39762, USA}
\affiliation{Kavli Institute of Theoretical Physics, University of California, Santa Barbara, Santa Barbara, CA 93106, USA}
\author{Yu Yi-Xiang}
\affiliation{Beijing National Laboratory for Condensed Matter Physics,
Institute of Physics, Chinese Academy of Sciences, Beijing 100190, China}
\affiliation{Beijing Institute of Science and Technology, Beijing 100083, China}
\author{Jinwu Ye}
\affiliation{Institute for Theoretical Sciences,
Westlake University, Hangzhou, 310024, Zhejiang, China}
\affiliation{Tsung-Dao Lee Institute, Shanghai 200240, China}
\affiliation{Department of Physics and Astronomy, Mississippi State University, MS 39762, USA}
\affiliation{Kavli Institute of Theoretical Physics, University of California, Santa Barbara, Santa Barbara, CA 93106, USA}
\author{W.-M. Liu}
\affiliation{Beijing National Laboratory for Condensed Matter Physics, Institute of Physics, Chinese Academy of Sciences, Beijing 100190, China}
\date{\today }

\begin{abstract}
We investigate chaotic to integrable transition
in two types of hybrid SYK models which contain both $ q=4 $ SYK with interaction $ J $
and $ q=2 $ SYK with an interaction $ K $ in type-I
or $(q=2)^2$ SYK with an interaction $ \sqrt{K} $ in type-II.
These models include hybrid Majorana fermion, complex fermion and bosonic SYK.
For the Majorana fermion case, we discuss both $ N $ even and $ N $ odd case.
We make exact symmetry analysis on the possible symmetry class of both types of hybrid SYK in the 10 fold way
by Random Matrix Theory (RMT) and also work out the degeneracy of each energy levels.
We introduce a new universal ratio which is  the ratio of the next nearest neighbour (NNN)
energy level spacing to characterize the RMT.
We perform exact diagonalization to evaluate both the known NN ratio and the new NNN ratio,
then use both ratios to study Chaotic to Integrable transitions (CIT) in both types of hybrid SYK models.
Some preliminary results on possible quantum analog of Kolmogorov-Arnold-Moser (KAM) theorem and its dual version in the
quantum chaotic side are given.
We explore some intrinsic connections between the two complementary approaches to quantum chaos: the RMT and
the Lyapunov exponent by the $ 1/N $ expansion in the large $ N $ limit at a suitable temperature range.
We stress the crucial differences  between the quantum phase transition (QPT) characterized by
renormalization groups at $ N=\infty $, $ 1/N $ expansions at a finite $ N $ and
the CIT characterized by the RMT at a finite $ N $: the former focus on the ground state and its low energy excitations
( edge states in the Fock space ), the latter on excited states ( bulk states in the Fock space ).
We also discuss Eigenstate Thermalization hypothesis (ETH)'s power on a quantum chaotic bulk state and
its inability to encode the edge states. Comments on some previously related works are given.
Some future perspectives, especially the failure of the Zamoloddchikov's c-theorem in 1d CFT RG flow are outlined.
\end{abstract}

\maketitle



\section{Introduction}
\label{sec:intro}

In classical chaos, the Lyapunov exponent was used to characterize the
exponential growth of two classical trajectories
when there are just a tiny difference in the initial conditions.
The classical concept of Lyapunov exponent can be extended to its quantum analog to characterize the exponential growth of two initially
commuting operators in the early time (up to the Ehrenfest time)
under the evolution of a quantum chaotic Hamiltonian
\cite{larkin,chaosbutter,chaosbound}.
There are recent flurry of research activities to extract the Lyapunov exponent
$\lambda_L$ of the Sachdev-Ye-Kitaev (SYK) model and
its various variants  through evaluating out of time ordered correlation (OTOC)
functions \cite{SY,SY3,Kit,subir3,Pol,Mald,CSYKnum,Gross,longtime1,longtime2,liu1,liu2}.
The quantum analog of $\lambda_L$  need to be evaluated away
from the thermodynamic limit by a $1/N $ expansion in the SYK models,
also away from the conformal invariant limit due to a leading irrelevant operator
and at a finite temperature (ranging from low to infinite temperatures).
There are also calculations in the dual bulk quantum black holes in an asymptotic $ AdS_2 $ geometry
to demonstrate the correspondence between the SYK models and  Jackiw-Teitelboim gravity \cite{Ads2Mald,JT1,JT2,JT3}.
The quantum chaos in the SYK models are due to the quenched disorders and interactions.
However, it inspired a new class of clean models called
(colored or un-colored) Tensor (Gurau-Witten) model
\cite{tensor1,tensor2,tensor3},
which share similar quantum chaotic properties
as the SYK models at least in the large $N$ limit.

From a completely different perspective and also at a much longer time scale ( Heisenberg time ),
the quantum chaos can also be characterized by the system's energy level-level correlations encoded in
the energy level statistics (ELS) and  the spectral form factor (SFF) through
random matrix theory (RMT)\cite{book,cite2,cite3}.
The ELS and SFF are always evaluated in a finite but sufficiently large system \cite{MBLSPT,randomM0,randomM,randomsusy1,randomsusy2}.
The ELS of SYK can be described by the Wigner-Dyson (WD) distributions
in a $N\pmod 8$ way \cite{MBLSPT,randomM0,randomM,10-fold1}:
$ N=2,6 $ Gaussian unitary ensembles (GUE),
$ N=0 $ Gaussian orthogonal ensemble (GOE),
$ N=4 $ Gaussian  symplectic ensemble (GSE).



Here we investigate chaotic to integrable transitions (CIT) in two types of hybrid SYK models.
The type-I  contains both the chaotic $q=4$ SYK with interaction $J$
and the integrable $q=2$ SYK with interaction $ K $.  It violates the particle-hole (PH) symmetry and
includes the hybrid Majorana fermion, complex fermion and hard-core bosonic SYK.
The type-II includes $ q=4 $ SYK  of either Majorana or complex fermion with interaction $ J $ and
$ (q=2)^2 $ SYK with the interaction $ \sqrt{K} $. It preserves the particle-hole (PH) symmetry.
In this work, we mainly use the RMT approach, but will also explore
some possible connections between the RMT approach and  the Lyapunov exponent which can be extracted from 
the out of time correlation functions (OTOC).

There are direct motivations to study both types of Hybrid SYK models. It was also known that the $ q=4 $ SYK model provides a concrete model to
realize Eigen-state Thermalization Hypothesis (ETH) \cite{ETHrev}.
Even the entanglement entropy of its ground state satisfies the
volume law instead of the more common area law.
Here it is important to study how the gapless quantum spin liquid (QSL) responses under the Type I or Type II kinds of
integrable  perturbations. It would also be interesting to study how
the ETH breaks down under the Type I or Type II kind of integrable  perturbations.
On the dual bulk gravity side, these kind of perturbations are also needed to probe the interior
behind the black hole horizon \cite{pure}.
Most importantly, in any possible experimental systems to realize SYK models,
the Type I and Type II terms are most common perturbations. As shown in Fig.\ref{sykrg} in the conclusion section,
the two types hybrid SYK models also provide specific examples of RG flow between two CFT$_1$ fixed points which violate
Zamoloddchikov's c-theorem.
When comparing with some previously studied hybrid models, one can see there are also many other motivations to study both types of hybrid SYK models. For example, the multi-channel Kondo models show non-Fermi liquid (NFL) behaviours,
so it is important to study how the NFL changes under spin or channel anisotropy \cite{kondo1,kondo2,kondo3,kondo4}.
The Kitaev honeycomb lattice model \cite{honey} hosts a Majorana fermion quantum spin liquid (QSL),
so it is important to study how the Majorana fermion QSL changes under adding other interactions such as a Heisenberg interaction  or a Dzyaloshinskii-Moriya (DM) term \cite{rh}.

We first introduce a new universal ratio which is the ratio of
next nearest neighbor (NNN) energy spacing to characterize the RMT classes
and also establish its relations with the well-known ratio of
the nearest neighbor (NN) energy spacing \cite{ratiosta}.
When a doubly degenerate level is split by a small perturbation,
the NN ratio is nearly zero and also rapidly changing,
the new NNN ratio can be used to  effectively characterize
the ``hidden'' RMT behaviours, especially the CIT.
We make exact symmetry analysis to classify the possible symmetry class
in the 10-fold way\cite{10-fold1,10-fold3,10-fold4},
and then perform exact diagonalization (ED) on all these hybrid models.
In the Majorana fermion case, we pay special efforts to classify the odd $ N $ case.
Both the NN ratio and the NNN ratio are evaluated
when a doubly degenerate level is split by a small perturbation.
In the type-I hybrid SYK, as $ K/J $ increases,
there is always a chaotic to integrable transition (CIT) from GUE to Poisson
in all the hybrid  Majorana or complex fermionic models, but not the hybrid bosonic model.
Starting from the type-I hybrid  $ q=4 $ Majorana or complex fermion SYK side, for the GOE case,
there is a GOE to GUE crossover first, then a CIT from the GUE to the Poisson;
for the GUE case,
there is a direct CIT from the GUE to the Poisson;
for the GSE case,
any small $K$ destroys the double degeneracy of the GSE,
the NN ratio rises to GUE, then a CIT from the GUE to the Poisson.
In this case,
the NNN ratio can be most effectively applied to describe the stability regime of GSE
near the chaotic side and may also be used to describe the whole crossover
until to the integrable side.

In the type-II hybrid SYK, the symmetry analysis alone
may not be able to distinguish between the chaotic $ q=4 $ SYK  and integrable  $(q=2)^2$ SYK.
As $ K/J $ increases, there are always CIT from the corresponding WD distribution
dictated by the symmetry of the  $ q=4 $ SYK to the Poisson controlled by the  $ (q=2)^2 $ SYK.
When there is a double degeneracy at the $ ( q=2 )^2 $ side,
the NNN can be most effectively applied to describe the CIT
and may also be used to describe the whole crossover until to the quantum chaotic side.

In classical chaos, the classical Kolmogorov-Arnold-Moser (KAM) theorem states
that if an integrable Hamiltonian $H_0$ is disturbed by a small perturbation $\Delta H$,
which makes the total Hamiltonian $H=H_0+\Delta H$ non-integrable.
If the two conditions are satisfied:
(a) $\Delta H$ is sufficiently small
(b) the frequencies $\omega_i$ of $H_0$ are in-commensurate,
then the system remains quasi-integrable.
Despite some previous efforts, the quantum analogue of the KAM theorem remains elusive.
Just like the quantum analog of Lyapunov exponent can be studied in the context of the SYK models \cite{SY,SY3,Kit,subir3,Pol,Mald,CSYKnum,Gross,longtime1,longtime2,Ads2Mald,liu1,liu2},
it is important to explore the quantum analog of KAM theorem in both types of hybrid SYK models.
We first give a definition of the quantum analogue of the KAM theorem and also its possible
dual form which is the stability of quantum chaos in the context of RMT. Then
we use both NN and NNN ratio to numerically characterize the KAM regime in the integrable side and
the stability of quantum chaos  in the chaotic side.
We give some preliminary results on possible quantum analog of the KAM theorem,
the stability of quantum chaos ( can also be called the dual form of the KAM theorem ),
especially its dependence on the system's size $ N $ in the two types hybrid SYK models.
By pushing the methods developed in \cite{GangTian} further, we leave rigorous mathematical treatments to a future publication \cite{math}.

We also stress the crucial differences between  the onset of quantum phase transitions (QPT) at $ N=\infty $ characterized
by renormalization group (RG), $1/N $ expansions and the onset of CIT characterized by the RMT at a finite $ N $.
The former is focusing on the changes in the ground state and low energy excitations
( called edge excitations in the Fock space ), there is an associated divergent length scale and
quantum critical scalings and a finite size scaling at a finite size.
While the latter is focusing on the changes in the correlations in the bulk energy levels in the Fock space,
there is no associated divergent length scale,
therefore no quantum critical scalings and no finite size scaling at a finite size.
We also discuss Eigenstate Thermalization hypothesis (ETH)'s power on a quantum chaotic bulk state and
its inability to encode the ground state and low energy excitations.
Some comments on previous works on type-I hybrid SYK models are given.
Some perspectives and possible future directions are outlined in the conclusion section.
Especially, we point out that the Zamoloddchikov's c-theorem of 2d CFT and its extensions to 2d boundary CFT  and higher dimensions break down in
1d CFT RG flow. This fact maybe related to $ NAdS_2/NCFT_1 $ may also be dramatically different than its high dimensional counterparts.

As mentioned in the first two paragraphs, there are two complementary approaches to describe the quantum chaos:
the OTOC at an early time and the RMT at a later time.
However, so far, it seems there is no clear connections between the two different ways to characterize the quantum chaos.
In this work,
we establish some intrinsic connections between the two approaches: the quantum chaos in the bulk energy levels described by the RMT implies the quantum information scramblings with a non-vanishing Lyapunov exponent $ \lambda_L $ by the OTOC in the edge levels
( the ground state and low energy levels ) in a suitable temperature range.

\section{A new ratio to describe RMT:
ratio of next nearest neighbor energy spacing }

In this section, we first review the known results on the statistics of
the NN energy level spacing $ ( r, \tilde{r} ) $,
then introduce a new ratio which is the NNN energy level spacing $ ( r', \tilde{r}' ) $,
then establish an approximate, but quite accurate relation between the two.

\subsection{Review on the ELS of NN energy level spacings}

Let $\{E_n\}$ be an ordered set of energy levels
and $s_n=E_{n+1}-E_{n}$ are the NN spacings.
By considering a $2\times2$ matrices system,
Wigner derived a simple approximate expression for the distribution function $P(s)$ of the NN spacing,
\begin{align}
    P_{w,\beta}(s)=a_\beta s^\beta e^{-b_\beta s^2}
\label{PW}
\end{align}
where $\beta=1,2,4$ is the Dyson index for GOE, GUE and GSE respectively.
It is also known that independent random energy levels would yield a Poisson distribution
\begin{align}
    P_{p}(s)=e^{-s}
\label{PP}
\end{align}
However, in order to compare different results from different systems,
the energy levels will need an unfolding procedure,
which is not convenient when large enough statistics is not available
To get rid of the dependence on the local density of states,
it is convenient to look at the distribution
of the ratio of two adjacent energy level spacings \cite{ratiosta,MBLSPT}
$r_{n}=s_{n}/s_{n+1}$ which distributes around $1$.
This quantity has the advantage that it requires no unfolding
since ratios of consecutive level spacings are independent of the local density of states.

By considering  $3\times3$ matrices system,
the authors in \cite{ratiosta} obtained the Wigner-like surmises of
the ratio of consecutive level spacings distribution
\begin{equation}
    P_{p}(r)=  \frac{1}{(1+r)^2},
    \quad
    P_{w}(r)=  \frac{1}{Z_{\beta}} \frac{(r+r^2)^{\beta} }{(1+r+r^{2})^{1+3 \beta/2}}
\end{equation}
where $\beta=1,2,4$
and $Z_{\beta}=8/27, 4 \pi/81\sqrt{3},  4 \pi/729\sqrt{3}$
for GOE, GUE and GSE respectively.
The distribution $P_W(r) $ has the same level repulsion at small $r$ as $P_W(s)$,
namely, $P_W(r)\sim r^\beta$.
However, the large $r$ asymptotic behavior $P_W(r)\sim r^{-(2+\beta)}$ is
dramatically different than the fast exponential decay of $P_W(s)$.

One may also compute the distribution of the logarithmic ratio
\cite{ratiosta,MBLSPT}  $P(\ln r)=P\left( r\right) r$.
Because $P \left( \ln r\right)dr $ is symmetric under $ r \leftrightarrow 1/r $,
one may confine $0<r<1$ and double the probability density
$P\left( \tilde{r}\right) =2P\left(r\right) $. Therefore, the above two distributions have two different sets of expected values of
$\tilde{r} = \min \{r, 1/r \} $:
\begin{align}
    \langle\tilde{r}\rangle_{p}
	&=\int_{0}^{1}2 r P_{p}(r) dr=2\ln 2-1 \approx 0.38629 ,    \nonumber   \\
    \langle\tilde{r}\rangle_{w}
	&=\int_{0}^{1}2 r P_{w,\beta=1,2,4}(r) dr
\end{align}%
which is $4-2\sqrt{3} \approx 0.53590$,
$2\sqrt{3}/\pi-1/2 \approx 0.60266$,
$32\sqrt{3}/(15\pi)-1/2 \approx 0.67617$
for GOE,GUE and GSE respectively.

\subsection{Introduction and calculation of the ELS of NNN energy level spacings }

When there are nearly double degenerated energy levels, we find that
it is convenient to introduce next nearest-neighbor (NNN) spacings
$s'_n=E_{2n+1}-E_{2n-1}$ and ratios $r'_n=s'_n/s'_{n-1} $.
Here we study the  distribution of the ratio of the two NNN spacings by exploring
the exact calculation for $5\times5$ matrices.

For a Poisson ensemble, it is more convenient to work with $P(s_1,s_2,s_3,s_4)$
and the ratio of consecutive NNN level spacings is $r'=(s_3+s_4)/(s_1+s_2)$.
The NNN ratio distribution can be calculated from
\begin{align}
    P_p^{(2)}(r')\!=\!\int \prod_{i=1}^4 ds_i
	P(s_1,s_2,s_3,s_4)\delta\Big(r'\!-\!\frac{s_3+s_4}{s_1+s_2}\Big)\>.
\end{align}
Since energy levels are not correlated in Poisson ensemble,
one can rewrite $P(s_1,s_2,s_3,s_4)=\prod_{i=1}^4P_P(s_i)$
and evaluate the integral to obtain a simple result
\begin{align}
    P_p^{(2)}(r')=\frac{6r'}{(1+r')^4}\>.
\label{eq:PPr'}
\end{align}
It is easy to see that $P_p^{(2)}(r')\sim r'$ when $r'$ is small,
and $P_p^{(2)}(r')\sim {r'}^{-3}$ when $r'$ is large.
Interestingly, there is a NNN level repulsion in Poisson ensemble
which intuitively can be understood due to
the interruption of the intermediate level.

\begin{figure}[!htb]
\includegraphics[width=0.9\linewidth]{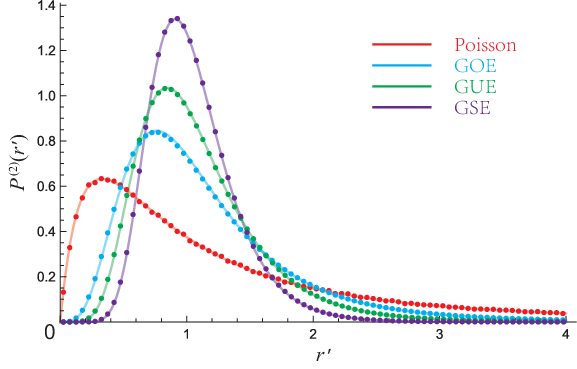}
\caption{Distribution of the ratio of consecutive
	 NNN level spacings $P^{(2)}(r')$ for Poisson and RMT ensembles:
	 solid lines are from exact evaluation of integral Eq.\eqref{eq:PWr'}.
	 The points are numerical data.
	 The Poisson data are obtained from $10^6$ independent generated random numbers.
	 The random matrix data are obtained by diagonalizing the corresponding
	 GOE (real), GUE (complex) and GSE (quaternion) matrices of
	 size $ N=1000 $ with Gaussian distributed entries,
	 averaged over $10^5$ histograms.}
\label{nnn1}
\end{figure}

For various Gaussian ensembles, it is good to start from
the joint probability distribution 
$\rho(E_1,E_2,E_3,E_4,E_5)$ and the ratio of consecutive NNN level spacings is
$r'=(E_5-E_3)/(E_3-E_1)$.
The NNN ratio distribution can be calculated from
\begin{align}
    P_{w,\beta}^{(2)}(r')\!=\!\!\int \prod_{i=1}^5 dE_i
	\rho(E_1,E_2,E_3,E_4,E_5)\delta\Big(r'\!-\frac{E_5-E_3}{E_3-E_1}\Big)
\end{align}
where the joint probability distribution takes the form
$\rho(E_1,E_2,E_3,E_4,E_5)
=C_{\beta} \prod\limits_{1 \leq i<j\leq5}|E_i-E_j|^\beta
\prod\limits_{i=1}^5 e^{-E_i^2/2}$.
The integral can be simplified as
\begin{align}
\label{eq:PWr'}
    P_{w,\beta}^{(2)}(r')
	&\sim  \int_0^\infty dx\int_0^{rx} dy\int_0^{x} dz  f(x,y,z,r') \>, \\
    f(x,y,z,r')
	&=e^{-\frac{1}{5}[(2+2r^{\prime 2}+r)x^2+2z^2+2y^2+yz+(1-r')x(y-z)]} \nonumber\\
	&\hspace{-1.8cm}
	\times x[r'(1\!+r')x^3(r'x\!-\!y)(r'x\!+\!z)y(x\!+\!y)(x\!-\!z)z(y\!+\!z)]^\beta ,   \nonumber
\end{align}
where $\sim$ means the normalization constant is ignored.
The integrals can be evaluated analytically,
but its expression is lengthy.
Here, we just show analytically its asymptotic behaviour:
$P_{w,\beta}^{(2)}(r')\sim {r'}^{3\beta+1}$ when $r'$ is small,
and $P_{w,\beta}^{(2)}(r')\sim {r'}^{-3(\beta+1)}$ when $r'$ is large.
If comparing them with those of $P_W(r)$,
one can see that the asymptotic behaviours of the NNN level statistics with index $\beta$
are the same as those of the NN  with index $3\beta+1$
(See Eq.\eqref{eq:PWW2} below for all the ranges).
Similar to the Poisson case discussed above,
the intermediate energy level induces much stronger level repulsions between NNN.
In Fig.\ref{nnn1}, we compare $P_{w,\beta}^{(2)}$ obtained from Eq.\eqref{eq:PWr'}
against the numerical simulations of the corresponding WD ensembles.
We find nearly perfect agreement in all ranges of $r'$.

It is easy to check that $P^{(2)}(r')$ have the same symmetry as $P(r)$, namely,
    $P^{(2)}(r')=\frac{1}{r^{\prime 2}}P^{(2)}(\frac{1}{r'})$,
thus we can still define a $\tilde{r}'$.
From the NNN ratio distribution given in Eq.\eqref{eq:PPr'} for Poisson and Eq.\eqref{eq:PWr'} for WD,
one can calculate expectation value for $r'$ and $\tilde{r}'$ as
\begin{align}
    \langle r'\rangle=\int_0^\infty dr' r'P(r'),\quad
    \langle \tilde{r}'\rangle=\int_0^1 dr' 2r'P(r')
\label{eq:r2}
\end{align}
which are listed in Table I.

\begin{table}[!htb]
\caption {List of numerical values of averages
$\langle r\rangle$ and $\langle \tilde{r}\rangle$
as well as $\langle r'\rangle$ and $\langle \tilde{r'}\rangle$
for various ensembles.
The values of $\langle r\rangle$ and $\langle \tilde{r}\rangle$
are taken from Ref.\cite{ratiosta}. Those of $\langle r'\rangle$ and $\langle \tilde{r}'\rangle$
are obtained from evaluation of Eq.\eqref{eq:r2}.
We also list the values of $\langle r'\rangle_{3\beta+1}$
and $\langle \tilde{r}'\rangle_{3\beta+1}$
by using the $\beta\to3\beta+1$ rule in Eq.\eqref{eq:PWW2}.}
\begin{tabular}{l*{4}{@{\hskip 0.25in}c}}
\hline\hline
Ensembles			& Poisson   & GOE & GUE & GSE  \\
\hline
$\langle r\rangle$		& $\infty$  & 1.75	& 1.3607    & 1.1747   \\
$\langle \tilde{r}\rangle$      & 0.38629   & 0.5359	& 0.6027    & 0.6762   \\ \hline
$\langle r'\rangle$		& 2	    & 1.1736	& 1.0972    & 1.0516   \\
$\langle\tilde{r}'\rangle$	& 0.5	    & 0.6769	& 0.7344    & 0.7910   \\  \hline
$\langle r'\rangle_{3\beta+1}$		&-- & 1.1747	& 1.0980    & 1.0681   \\
$\langle\tilde{r}'\rangle_{3\beta+1}$	&-- & 0.6762	& 0.7335    & 0.7672   \\
\hline\hline
\end{tabular}
\label{tab:NN_NNN}
\end{table}

\subsection{An approximate but accurate  relation between the ELS of NNN and those of NN }

In fact, instead of lengthy results from exact evaluation of integral,
we find an approximate relation between $P^{(2)}(r')$ and $P(r)$:
\begin{align}
    P_{w,\beta}^{(2)}(r)\approx P_{w,3\beta+1}(r)
\label{eq:PWW2}
\end{align}

\begin{figure}[!htb]
\centering
\includegraphics[width=0.9\linewidth]{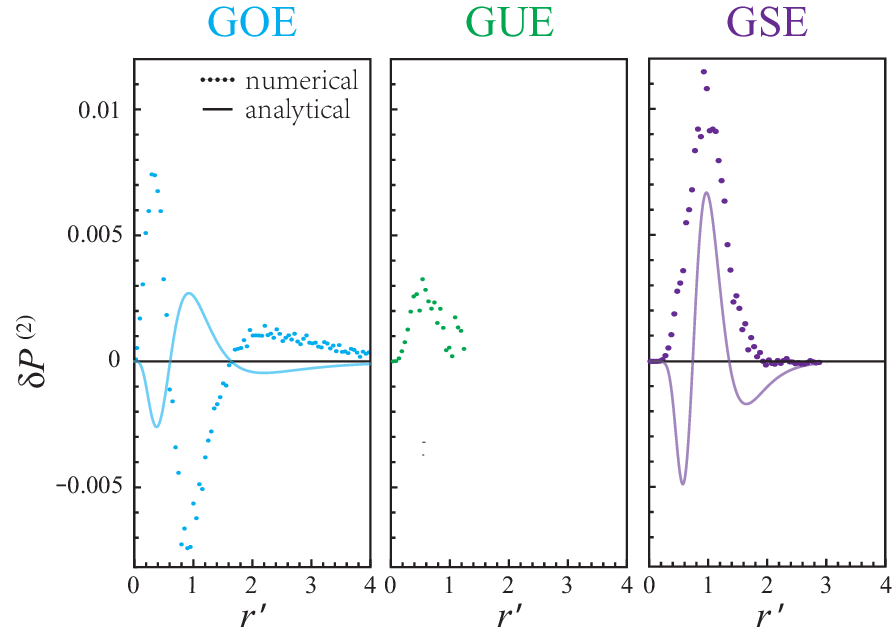}\quad
    \caption{ Difference of $\delta P^{(2)}=P_{W,\beta}^{(2)}(r)-P_{W,3\beta+1}(r)$.
    The solid line is from Eq.\ref{eq:PWr'}.
	The numerical data in $\delta P^{(2)}=P_\text{num}^{(2)}(r)-P_{W,3\beta+1}(r)$ are taken from Fig.\ref{nnn1}.
	Small $|\delta P^{(2)}|$ justifies validation of Eq.\eqref{eq:PWW2}}.
\label{nnn2}
\end{figure}

  The difference of the two was shown in Fig.\ref{nnn2}. The very small deviation
  shows that the approximation in Eq.\eqref{eq:PWW2} is quite accurate.
  In the last two lines in the TABLE I, we also list the numerical values of
  $ \langle r'\rangle_{3\beta+1} $ and  $\langle\tilde{r}'\rangle_{3\beta+1}$	
  using $ P_{W,3\beta+1}(r) $. All these values are very close to those using $ P_{W,\beta}^{(2)}(r) $ in Eq.\ref{eq:PWr'}.
  For example, just take $ \langle\tilde{r}'\rangle $ for the GSE case, one can see the relative difference
  $\frac{0.7672\, -0.7910}{0.7910}=-3\%$ is very small.

  In fact, one can see Eq.\ref{eq:PWW2} can also work well when putting $ \beta=0 $. Namely, when the NN satisfies Poisson, the NNN
  seems fit GOE  approximately.
For example, from the TABLE I,
$ \langle r'\rangle $ and $\langle\tilde{r}'\rangle$ for Poisson are $2$ and $0.5$
which are not too much away from
$\langle r\rangle $ and  $\langle \tilde{r}\rangle$ for GOE ( $1.75$ and $0.53590 $ ).
At least, one can use GOE for an quick eye guides to judge the
NNN for the Poisson as we did in all the following figures.

  In the following sections, we will apply both NN and NNN ELS to study the CIT
  in the two types of hybrid Sachdev-Ye-Kitaev models.

\section{Type-I Hybrid Sachdev-Ye-Kitaev models}

 By Type-I, we mean the integrable side is given by $ q=2 $ SYK which breaks the PH symmetry Eq.\ref{PH} of the $ q=4 $ SYK.
 We will discuss Majorana and complex type-I hybrid SYK respectively.

\subsection{The hybrid of $ q=2 $ and $ q=4 $  Majorana fermion SYK }

The CIT may be first investigated
in the hybrid of $ q=2 $ and $ q=4 $ Majorana fermion SYK,
which also known as type-I hybrid Majorana SYK model:
\begin{equation}
H_{M\text{-I}}=  \sum^{N}_{i<j<k<l} J_{ijkl} \chi_{i} \chi_{j} \chi_k \chi_l + i \sum^N_{i<j}
K_{ij} \chi_{i} \chi_{j}
\label{mix1}
\end{equation}
where  $  J_{ijkl}, K_{ij} $ are real and  satisfy the Gaussian distributions with
  $ \langle J_{ijkl} \rangle=0,  \langle J^2_{ijkl} \rangle= 3! J^2/N^3 $ and
  $ \langle K_{ij} \rangle=0,  \langle K^2_{ij} \rangle=  K^2/N $ respectively.
In the following,
we denote the first term of $H_{M\text{-I}}$, $ q=4 $ SYK model, as $H_{M4}$,
and the second term, $ q=2 $ SYK model, as $H_{M2}$.


In the $ J=0 $ limit, the $ q=2 $ SYK $H_{M2}$ breaks the PH symmetry Eq.\ref{PH}.
It is non-interacting, so must be integrable.
Its single particle density of state (DOS) satisfies the semi-circle law \cite{Gross}, while
its many body ELS satisfies the Poisson distribution \cite{Mald,Gross,randomM}.
Its many body low or high energy excitation level spacing is $ \sim 1/N $,
the low energy quasi-particle picture holds.
Its $ T=0 $ entropy density $ s_0$ vanishes.
This is very similar to the $ U(1)/Z_2 $ Dicke model
\cite{u1largen,gold,comment,strongED,KAM}
in the $ U(1) $ limit (inside the superradiant phase)
where the many body ELS also satisfies the Poisson distribution,
and the low energy excitation level spacing is also $ \sim D \sim 1/N $ where $ D $ is the
phase diffusion constant.

In the $ K=0 $ limit, the $ q=4 $ SYK Hamiltonian $H_{M4}$
is non-integrable at any finite $ N $.
In the following, we discuss when $ N $ is even or odd respectively.
For even number of sites $ N=2N_c $, one can split the sites to even and odd,
then introduce $ N_c $ complex fermions\cite{CSYKnum,MBLSPT,randomM}
by  $ c_i= ( \chi_{2i}-i \chi_{2i-1} )/\sqrt{2},
   c^{\dagger}_i= ( \chi_{2i}+i \chi_{2i-1} )/\sqrt{2} $
and define the PH symmetry operator to be \cite{evenodd}
\begin{equation}
   P= K \prod^{N_c}_{i=1} ( c^{\dagger}_i + c_i )
\label{PH}
\end{equation}
   It is easy to show $ P^2= (-1)^{[\frac{N_c}{2}]},  P c_i P= \eta c^{\dagger}_i, P c^{\dagger}_i P= \eta c_i,
   P \chi_i P= \eta \chi_i $
   where $ \eta=(-1)^{[\frac{N_c-1}{2}]} $. The total number of  fermions $ Q_c= \sum^{N_c}_{i=1} c^{\dagger}_i  c_i $.
   It is not a conserved quantity, but the parity $ (-1)^{Q_c} $ is in $H_{M4}$.
   Then  $ P Q_c P^{-1}= N_c-Q_c $ which justifies $ P $ as an anti-unitary PH transformation. $ P $ also commutes with the Hamiltonian \cite{TR}
$ [ P, H_{M4}] =0 $.
Depending on $ N \pmod 8 $, the ELS satisfies
GUE when $ N \pmod 8=2,6 $,
GOE when $ N \pmod 8=0 $,
GSE when $ N \pmod 8=4$  \cite{CSYKnum,MBLSPT,randomM}.
The ELS, the degeneracy at a given parity sector,
and the total parity sector are listed in the Table IIa.
Its low or high energy excitation level spacing is $ e^{-s_0 N} $
which leads to extensive $ T=0 $ entropy $ s_0=0.232424.... $
(in the  $ N  \rightarrow \infty $ limit before $ T \rightarrow 0 $ limit)
\cite{SY3,subir3,Mald}.
The system's many body DOS (different from the single particle DOS)
has been worked out  in \cite{Mald} to be similar, but different than the semi-circle DOS.
The quasi-particle picture completely breaks down.
This is similar to the $ U(1)/Z_2 $ Dicke model in the $ Z_2 $ limit
(inside the superradiant phase)
where the ELS satisfies the GOE distribution \cite{strongED,KAM} ( See Sec V-C ),
its two lowest energy splitting between two opposite parities is $ e^{-N} $
which is due to the instanton quantum tunneling (QT) process.
Of course, the ground state of $ q=4 $ SYK  is a gapless quantum spin liquid.
While that of  $ U(1)/Z_2 $ Dicke model is a gapped $ Z_2 $ symmetry broken super-radiant state
at $ N=\infty $.

\begin{figure}[!htb]
\centering
\includegraphics[width=\linewidth]{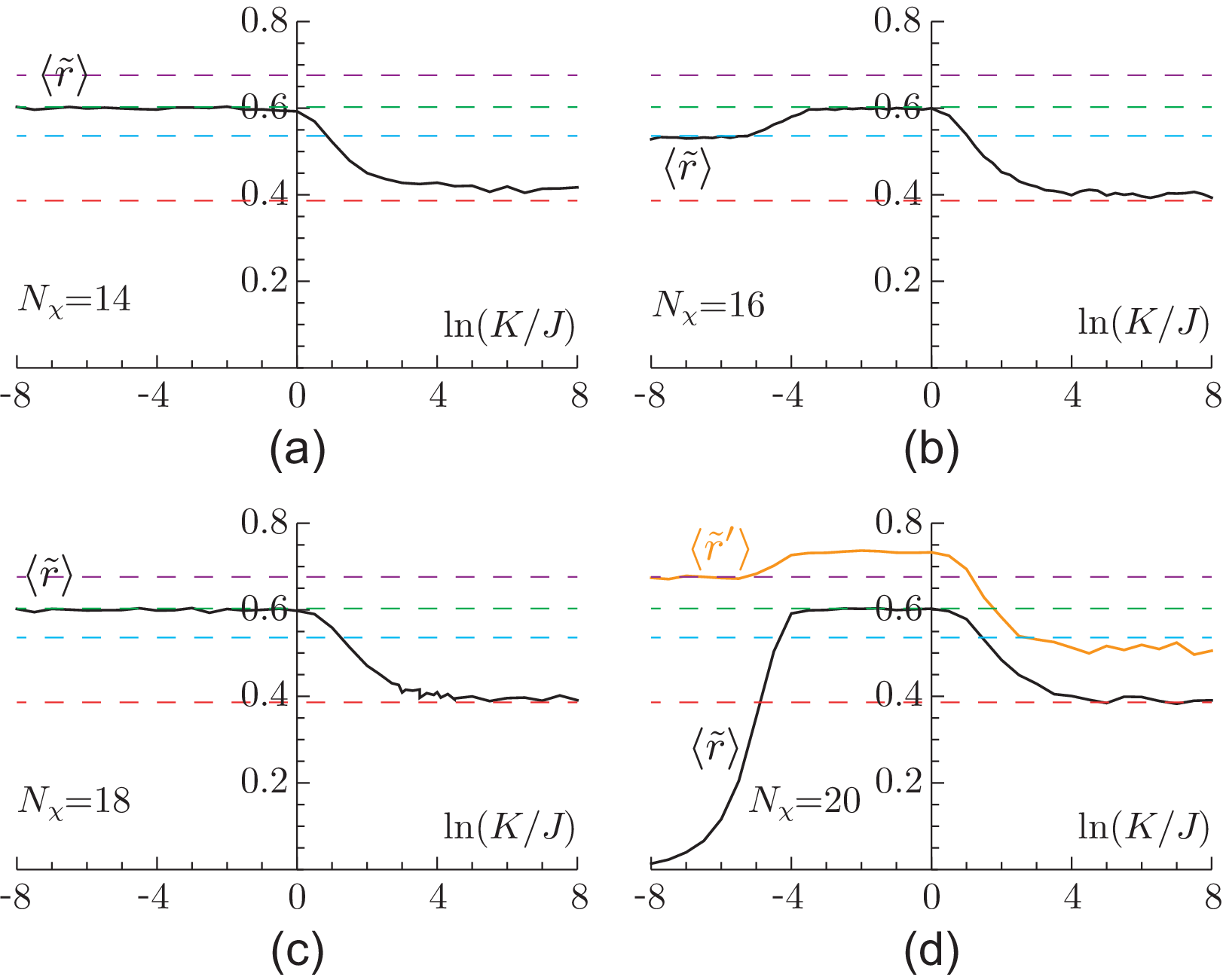}
\caption{  Even $N $: the evolution of the ELS for type-I hybrid Majorana SYK model with even $N $. (a) and (c) are GUE, (b) is GOE,
(d) is GSE on the $ q=4 $ side. $\langle\tilde{r}\rangle$ (black curve), and $\langle\tilde{r}'\rangle$ (orange curve) for the GSE case in (d)
	evaluated for  $N_\chi=14,16,18,20$ and are averaged over  $100$, $80$, $60$, $40$ samplings  respectively.
Notably, in the GSE case in (d), the NN ratio $ \langle \tilde{r}  \rangle $ is rapidly changing near the $ q=4 $ side, so it
is quite difficult to determine the stability regime of the quantum chaos. Fortunately,
the NNN ratio $ \langle \tilde{r}'  \rangle $ shows a nice plateau regime near the $ q=4 $ side, the quantum chaos stability regime can be
easily identified and listed in the Table III-VI. This dramatic advantage of the new  NNN ratio over the known NN ratio when there is a double
degneracy was
further demonstrated in all the following relevant figures when there are double degeneracy on the chaotic side or integrable side. 	}
\label{MajoranaSYK_Type1_Even}
\end{figure}

\begin{figure}[!htb]
\centering
\includegraphics[width=\linewidth]{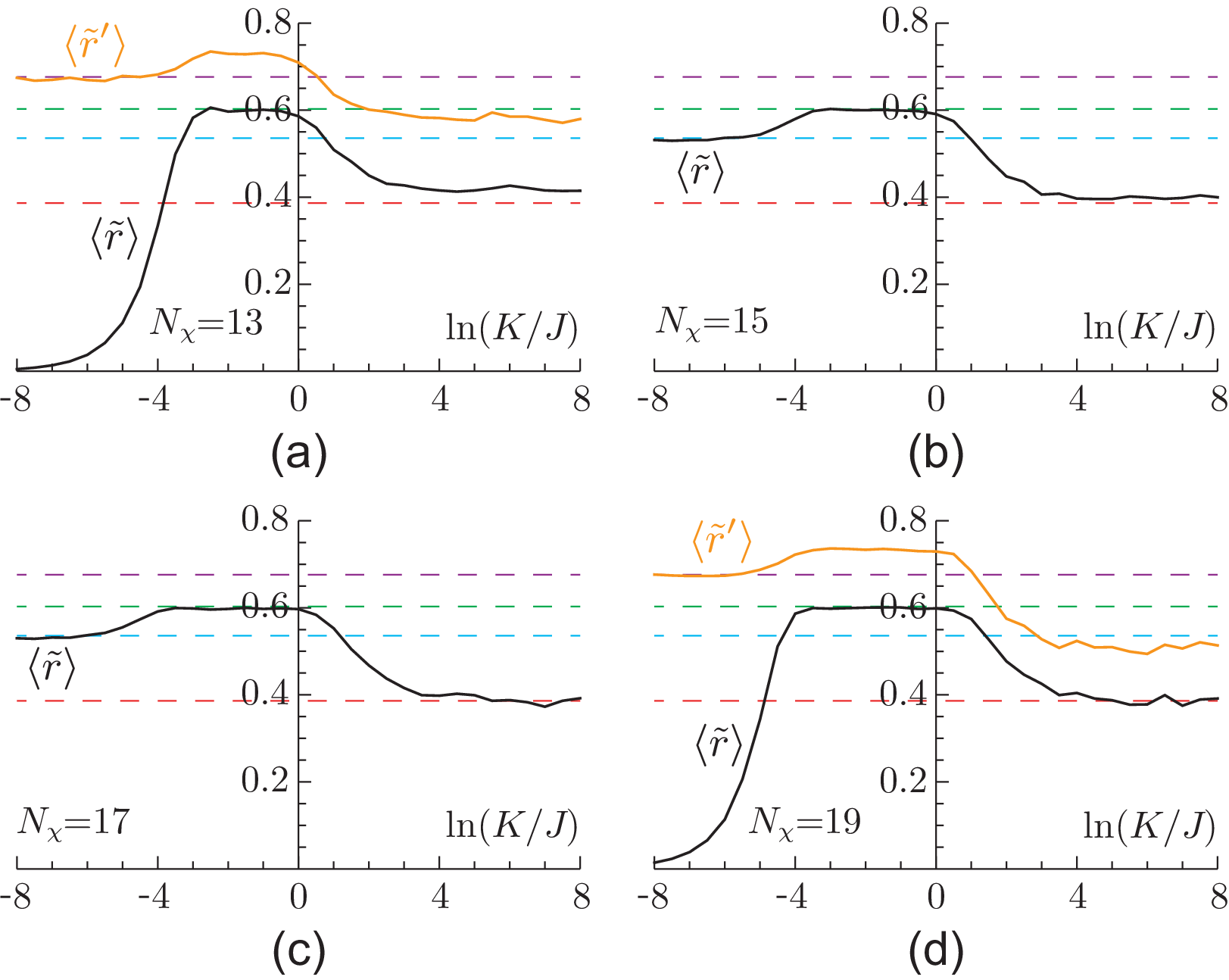}
\caption{ Odd $ N $: the evolution of the ELS for type-I hybrid Majorana SYK Model with odd $N $.
(b) and (c) are GOE, (a) and (d) are GSE on the $ q=4 $ side. $\langle\tilde{r}\rangle$ (black curve),
	and $\langle\tilde{r}'\rangle$ (orange curve) for the GSE case in (a) and (d)
	evaluated for $N_\chi=13,15,17,19$ and are averaged over $100$, $80$, $60$, $40$ samplings respectively.
    Note the advantages of using $ \langle \tilde{r^{\prime}}  \rangle $ over $ \langle \tilde{r}  \rangle $,
    especially in the quantum chaos side, in (a) and (d). }
\label{MajoranaSYK_Type1_Odd}
\end{figure}


   However, when $N$ is odd, the above procedures for even $ N $ needs to be modified.
   In fact, one can still take the advantage of the above representation with $ N $ even case by
   adding $ \chi_{N+1} =\chi_{\infty} $  to make the parity conservation  explicitly, but also enlarge the Hilbert space twice.
   Similar strategy was used before to study the symmetry protected topological phase of
   odd number of Majorana chain \cite{kitSPT} and the ELS of the SYK model with $ N $ odd \cite{MBLSPT,unknown}.
   Then one can still define $ P $ with $ N_c=\frac{N+1}{2} $ as before.
   All the commutation relations still apply.

   So when $N\pmod 8=3,7$, $N_c$ is even, then $P^2=-1,+1$ respectively.
   Under $P$,  $Q_c$ maps to the same parity sector.
   So it is in GSE and GOE with degeneracy $d=2$ and $d=1$ respectively at a given parity sector.
   When using the $Z$ operator (see Eq.\eqref{z} below) which maps
   $ Q_c $  to $ Q_c + 1 $, so it establishes the connection between the two parity sectors.
   So it has the degeneracy $ d_t=2+2 $ and $ d_t=1+1 $  when considering both parities.

   When $N \pmod 8=1,5$, $ N_c $ is odd, then $ P^2=+1,-1 $ respectively.
   However, under $ P $,  $ Q_c $ maps to the opposite sector $ Q_c +1 $.
   So one may only use $ P $ to establish the connection between the two opposite parity sectors.
   This forced us to look for a new operator which may map $  Q_c $
   into the same sector and still commutes with the Hamiltonian. This new operator is found to be:
\begin{eqnarray}
 Z  =  P \chi_{\infty}= K \prod^{N_c-1}_{i=1} ( c^{\dagger}_{i} + c_{i} )
\label{z}
\end{eqnarray}
  which can be written down by just changing $ N_c $ in $ P $ to $ N_c-1 $.
  So it will play a complementary role as $ P $ which will be analyzed in the following.

  One can show that $  Z \chi_{i} Z  = (-1)^{N_c} \eta \chi_{i} $
  where  $ i=\infty $ is  excluded and the number $  \eta=(-1)^{[\frac{N_c-1}{2}]}  $ is defined below Eq.\ref{PH}.
 It is also easy to see that  $ Z^2=\eta $. Using the fact that the Hamiltonian does not contain the
 fermions added at infinity, one can show that it still commutes with the Hamiltonian $ [Z,H_{M4}]=0 $.
 It also leads to $ Z Q_c Z^{-1}= N_c-1-Q_c + 2 n_{ \infty} $  where $ n_{ \infty}= c^{\dagger}_{\infty} c_{\infty}=
 \frac{1}{2}- i \chi_{ \infty} \chi_{N}$.

 So when $N \pmod 8=1,5$,  $ N_c $ is odd. Under $ Z $, $ Q_c $ maps to the same sector.
 $ Z^2=1 $ and $ Z^2=-1 $ respectively, so it is in GOE and GSE
 with degeneracy $ d=1 $ and $ d=2 $ respectively at a given parity sector.
 When using the $ P $ operator (see above) which maps
 $ Q_c $  to $ Q_c + 1 $, so it establishes the connection between the two parity sectors.
 Then it has the degeneracy $ d_t=1+1 $ and $ d_t=2+2 $   when considering both parities.

\begin{table}[!htb]
\caption{
The ELS and degeneracy of the Majorana fermion $ q=4 $ SYK model at N even in (a) or odd in (b).
The degeneracy $ d $ is at a given parity sector $ (-1)^{ Q_c } $.
The total degeneracy $ d_t $ is at both parity sectors.
(a) When $N$ is even, there is only one anti-unitary operator $ P $.
When $N \pmod 8=2,6$, $ P $ maps $ Q_c $ into $  Q_c + 1 $.
When $N \pmod 8=0,4$, $ Q_c $ and $ Q_c + 1$ are completely dependent, no operator connects between the two opposite parities.
(b) When $N$ is odd, after adding one Majorana fermion at $ N+1=\infty $, one doubles the Hilbert space,
also introduces one more conserved quantity ( the parity ),
there are two anti-unitary operators $ P $ and $ Z $.
When $N \pmod 8=3, 7$,  $ P $ maps $ Q_c $ to itself, $ Z $ maps $ Q_c $ into $  Q_c + 1 $.
When $N \pmod 8=1, 5$, $ Z $ maps $ Q_c $ to itself, $ P $ maps $ Q_c $ into $  Q_c + 1 $.
So $ P $ and $ Z $ exchange their roles in the two cases. So the $ d_t $ in the odd N case
is the degeneracy in the enlarged Hilbert space which is the twice of the original one \cite{unknown}. }
\begin{tabular}{ |*{5}{c|}  }
\hline\hline
$N\pmod 8$  & 0		& 2	    & 4		& 6	    \\  \hline
ELS         & GOE	& GUE	    & GSE	& GUE	    \\  \hline
$ \beta $   & 1		& 2	    & 4		& 2	    \\  \hline
$ Q $       & $d=1$	& $d=1$	    & $d=2$	& $d=1$	    \\  \hline
$ Q_t $	    & $d_t=1$   & $d_t=1+1$ & $d_t=2$   & $d_t=1+1$ \\   \hline
\hline
$N\pmod 8$  & 1		& 3	    & 5		& 7	    \\  \hline
ELS         & GOE	& GSE	    & GSE	& GOE	    \\  \hline
$ \beta $   & 1		& 4	    & 4		& 1	    \\  \hline
$ Q $       & $d=1$	& $d=2$	    & $d=2$	& $d=1$	    \\  \hline
$ Q_t $	    & $d_t=1+1$ & $d_t=2+2$ & $d_t=2+2$ & $d_t=1+1$   \\
\hline\hline
\end{tabular}
\end{table}

   Now we apply the PH transformation to the hybrid SYK model Eq.\ref{mix1}.
   The parity  $ (-1)^{Q_c} $ remains to be conserved.
   However, $ P $ ( when $ N \pmod 8=1, 5 $, one use  $ Z $, in all the other cases, one use $ P $  )
   is not conserved anymore due to $ \{ P, H_2 \} =0 $ (or $ \{ Z, H_2 \} =0 $).
   So the hybrid SYK does not have the PH symmetry anymore.
   Just from symmetry point of view (the 10-fold way classification scheme),
   the hybrid Majorana SYK Eq.\ref{mix1} belongs to the class A,
   so may satisfy GUE for any ratio of $ K/J $.
   Our ED studies \cite{JW} at a given parity $ (-1)^{Q_c} $ sector  were shown in
   Fig.\ref{MajoranaSYK_Type1_Even} for even $ N $ and \ref{MajoranaSYK_Type1_Odd} for odd $ N $.
   In the following, we will discuss them respectively.

   For even $ N $, there are 3 cases:
    (a) For $ N \pmod 8=0 $, the $ q=4 $ Majorana fermion SYK at $ K=0 $ is in GOE,
    the  hybrid is in the GUE around $ K/J =e^{-3.5} \sim 0.03 $ to $1$,
    there is a crossover from GOE to GUE first,
    then a CIT from GUE to the Poisson as $ K/J $ increases.

    (b) For $ N \pmod 8=2,6 $, the $ q=4 $ Majorana fermion SYK at $ K=0 $ is in GUE,
    it stays in the GUE until $ K/J =1$,
    then there is a CIT from GUE to the Poisson as $ K/J $ increases.

    (c) For $ N \pmod 8=4 $,  the $ q=4 $ Majorana fermion SYK at $ K=0 $ is in GSE at $ K=0 $.
    Because $ P^2=-1 $, any energy level is doubly degenerate at a given parity sector,
    so when doing ELS, we only pick up one of the  doubly degenerate levels to demonstrate the GSE.
    Any small $ K $ breaks the degeneracy, then we may consider both sets of energy levels,
    a small $ K $ makes $ \langle \tilde{r} \rangle $ small,
    so $ \langle \tilde{r} \rangle $ starts from zero and increases as $ K/J $ increases,
    then reaches the GUE in the range from $ e^{-4} $ to $ e^{0.5}$.
    There is a CIT from GUE to the Poisson as $ K/J $ increases.
    So in this case, using the NN ELS is not enough.
    One may start to use the combination of NN and the NNN ELS presented in Sec.2.
    The $ \langle\tilde{r}'\rangle $ was also shown in
    Fig.\ref{MajoranaSYK_Type1_Even}d.
    It is convenient to combine both $\langle \tilde{r}\rangle$ and $ \langle\tilde{r}'\rangle $ into the same Fig.3d.
    When $\langle \tilde{r}\rangle$ in (d) is close to be zero,
    the NNN $ \langle\tilde{r}'\rangle $ in (d) still shows GSE until $ K/J \sim e^{-5} $.
    When $\langle \tilde{r}\rangle$ in (d) reaches the plateau value
    $ \sim 0.60266 $ of GUE with $ \beta=2 $, then according to Eq.\ref{eq:PWW2},
    $ \langle\tilde{r}'\rangle $ in (d) reaches the corresponding plateau value
    listed in Table I as $ \sim 0.7344 $ with a RMT index $ 3 \beta + 1=7 $.
    When $\langle \tilde{r}\rangle$ in (d) reaches the plateau value
    $ \sim 0.38629 $ of Poisson, according to Eq.\ref{eq:PWW2},
    $ \langle\tilde{r}'\rangle $ in (d) reaches a corresponding plateau value
    listed in Table I as $ \sim 0.5 $.
    As pointed out in Sec.2, it is only slightly below the GOE value $ 0.53590 $
    with the RMT index $ 3 \beta + 1=1 $.

     For odd $ N $,  due to the absence of the GUE in Table IIb, at the $ q=4 $ side,  there are only
     the GSE case in (a) and (d), the GOE case in (b) and (c).
     They show similar evolution patterns as the corresponding GSE and GOE cases
     at the $ q=4 $ side for even $ N $ cases shown in Fig.\ref{MajoranaSYK_Type1_Even}.


   Obviously, at any given disorder realization of $ K_{ij} $, the eigen-energies
   of $ q=2 $ SYK is in-commensurate (so the disorder in SYK may play a similar role
   as the Berry phase in the $ J-U(1)/Z_2 $ Dicke model),
   a quantum version of Kolmogorov-Arnold-Moser ( KAM ) theorem \cite{KAM} ( See Sec V-C ) should apply when $ J/K $ is sufficiently  small ( See Sec V-C ).
   So the ELS changes from the Poisson to GUE  around some critical $ (J/K)_c $ values.
   The mean value $ \langle \tilde{r} \rangle $ also changes from its Poisson value to the corresponding GUE value.
   It is similar to the $ \beta_{c} $ in the $ U(1)/Z_2 $ Dicke model where the
   ELS changes from  the Poisson to GOE when inside the superradiant phase $ g/g_c \geq 1 $ (see Fig.2a in \cite{KAM}).
   Of course, in contrast to the Dicke model, due to the quenched disorders,
   there is no regular regime (see Fig.2b in \cite{KAM})
   at any values of $ J/K $.  Compared to the ED to study the CIT in the $ U(1)/Z_2 $ Dicke model, the extra work needed  here is that one need to draw the random couplings
 $ J_{ijkl}, K_{ij} $ from the corresponding Gaussian distributions $ P[J] $ and $ P[K] $,
 then get ELS at a given set of $ J_{ijkl}, K_{ij} $.
 Then one need to repeat the same calculations over 40 to 100 samples of such a random realizations of $ J_{ij,kl}, K_{ij} $,
 then  perform the average of ELS over these 40 to 100 samples \cite{noise}.

   One may also understand the quantum analog of the KAM theorem from a dual point of view, namely,
   the stability of quantum chaos of
   the quantum chaotic $ q=4 $ SYK against the non-chaotic perturbation as one turns on $ K/J $.
   The dual form of the KAM theorem states that as $ K/J $ increases, there should be a crossover
   from the WD to the GUE, then a CIT from the GUE to the Poisson. Our ED studies shown in
   Fig.\ref{MajoranaSYK_Type1_Even} and \ref{MajoranaSYK_Type1_Odd} confirms this  global picture.


\subsection{The hybrid of $ q=2 $ and $ q=4 $  complex fermion SYK }

    The Majorana fermion SYK was extended to the complex fermion which has a $ U(1) $
    charge symmetry  \cite{CSYKnum}.
    One may also add a chemical potential $ \mu $ to fix the conserved
    fermion filling factor \cite{can}:
\begin{equation}
     q_c= \sum_{i} ( c^{\dagger}_i c_i-1/2)
\label{qc}
\end{equation}
    The procedures used for the Majorana fermion in the last subsection
    ( of course, no odd $ N $ case anymore ) can also be applied
    to study the $ q=2 $ and $ q=4 $ type-I hybrid complex SYK:
\begin{equation}
H_{C}=  \sum^{N}_{i<j,k<l} J_{ij;kl} c^{\dagger}_{i} c^{\dagger}_{j} c_k c_l +  \sum^N_{i<j}
K_{ij} c^{\dagger}_{i} c_{j}- \mu q_c
\label{mix2}
\end{equation}
    where  $ \langle J_{ij;kl} \rangle=0,  \langle | J_{ij;kl} |^2 \rangle= 3! J^2/N^3 $ drawing from
  the Gaussian distribution $ P[J_{ij;kl}] \sim e^{ - A  |J_{ij,kl}|^{2}/2 J^2 } $ where $ A= N^3/3 ! $.
  In general, $ J_{ij;kl}=-J_{ji;kl}, J_{ij;kl}=-J_{ij;lk}, J^{*}_{ij;kl}= J_{kl;ij} $.
 {\sl  We also take the four site indices $ i,j; k,l$ are all different \cite{matter} to keep the PH symmetry explicit at $ \mu=0 $.}
    $ K^{*}_{ij} =K_{ji} $ is  a Hermitian matrix satisfying
    $ \langle K_{ij} \rangle=0,  \langle |K_{ij}|^2 \rangle=  K^2/N $.

\begin{figure}[!htb]
\centering
\includegraphics[width=\linewidth]{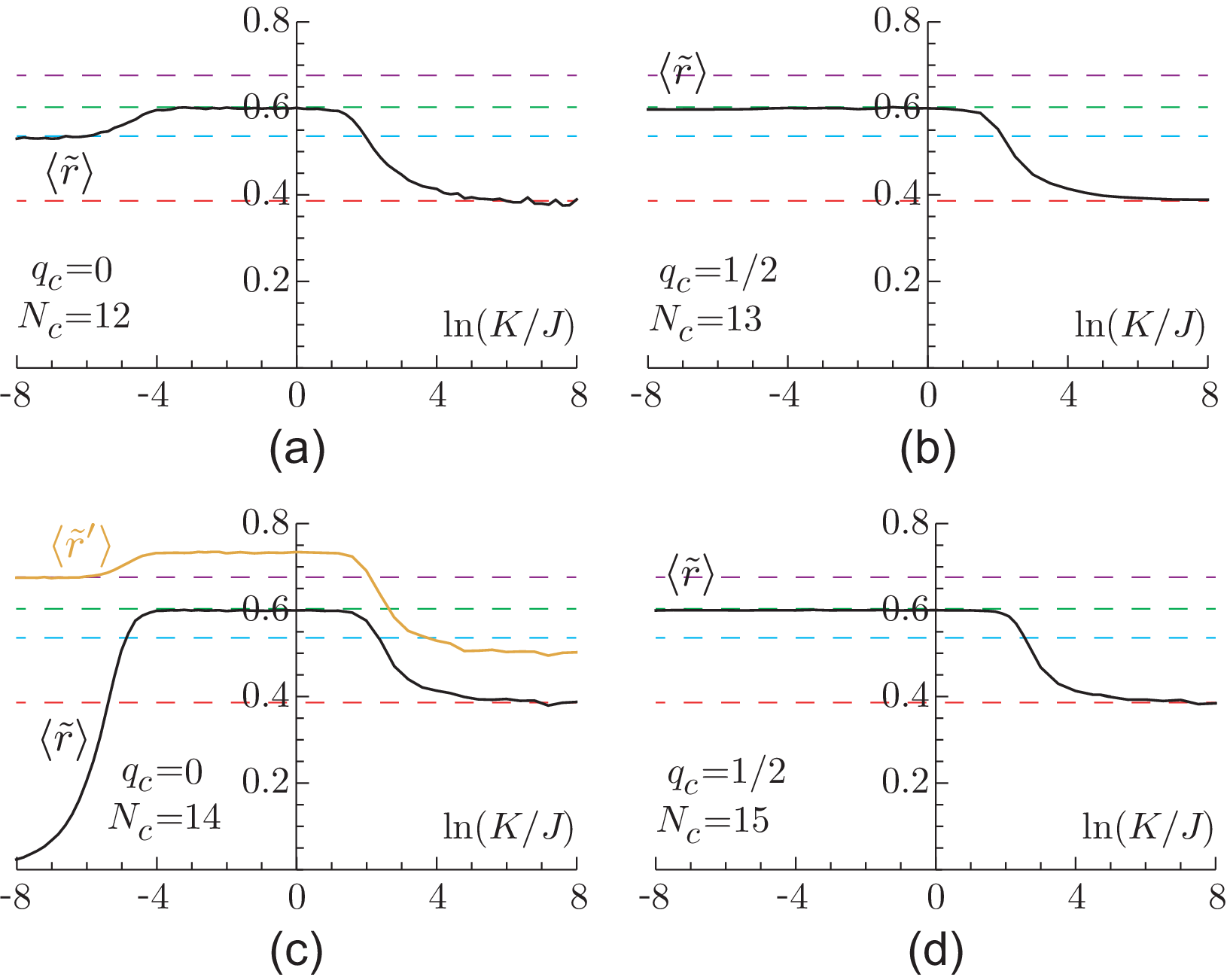}
\caption{ The evolution of the ELS for type-I hybrid complex SYK model. (a) is GOE
(b) and (d) are GUE, (c) is GSE on the $ q=4 $ side.
$\langle\tilde{r}\rangle$ (black curve),
	and $\langle\tilde{r}'\rangle$ (orange curve) for GSE case in (c)
	evaluated at $N_c=12,13,14,15$ and are averaged over
	$100$, $50$, $50$, $50$ samplings respectively.
Note the advantages of using $ \langle \tilde{r}' \rangle $
over $ \langle \tilde{r}  \rangle $ in (c), especially in the quantum chaos side.}
\label{ComplexSYK_Type1}
\end{figure}

 In the $ K/J=0 $ limit, the $ q=4 $ complex SYK is non-integrable at any finite $ N $.
 Under the PH transformation, $ q_c \rightarrow -q_c $.
 So $ q_c=0 $ is the PH symmetric and only happens when $ N $ is even.
 At the half-filling $ q_c=0 $, the system also has the maximum zero temperature entropy $ s_0 $ \cite{CSYKnum}.
 Away from the half-filling $ \mu \neq 0 $, it breaks the $ P $ symmetry,
 $ -N/2 < q_c \neq 0 < N/2 $  corresponds to a non-vanishing electric field
 in a charged black hole in an asymptotic $ AdS_2 $ bulk \cite{subir3}.
    It was shown in \cite{MBLSPT} that when $ N \pmod 4=0,2 $, the ELS is GOE and GSE respectively.
    But when $ N \pmod 4=1,3 $ and $ q_c=\pm 1/2 $, it is GUE.
   In fact, as long as $ q_c \neq 0 $,  there is no PH symmetry anymore, so it is in GUE regardless of the value of $ N \pmod 4$.

   Now we apply the PH transformation to the hybrid SYK model Eq.\ref{mix2} when $ K/J \neq 0 $.
   The fermion number remains to be conserved.
   However, $ P $ is not conserved anymore  due to $ \{ P, H_2 \} =0 $
   ( here and thereafter, $ H_2 $ excludes  $ \mu $ in Eq.\ref{mix2} ).
   So the hybrid SYK does not have the PH symmetry anymore.
   Just from symmetry point of view, the hybrid complex SYK belongs to the class A in the 10 fold way classification,
   so may satisfy GUE for any ratio of $ K/J $.
   Our ED studies \cite{JW} at a given $q_c $ in Fig.\ref{ComplexSYK_Type1} shows
   that this is true only in the intermediate regimes of $ K/J $.
   The KAM theorem and its dual form applies in the two end regimes $ K/J \ll 1 $ where the ELS remains Poisson
   and $ K/J \gg 1 $ where the ELS becomes the WD determined by the $ q=4 $ SYK respectively.

 In our ED studies \cite{JW},
 we first study $ q=4 $ SYK to reproduce the known results in \cite{MBLSPT},
 then look at the $ q=2 $ SYK to show that it indeed satisfies the Poisson distribution.
 Then we study the hybrid model Eq. \ref{mix2} at a given $q_c $ at any $ K/J $.
 The ED result is shown in Fig.\ref{ComplexSYK_Type1} where there are also 3 cases:

(a) For $ N_c \pmod 4=0 $, at the half filling $ q_c=0 $,
the hybrid complex fermion SYK is in the GUE in a wide range near $ K/J =1 $,
there is a crossover from GOE to GUE, then a CIT from GUE to the Poisson as $ K/J $ increases.

   (b) Away from the half filling $ q_c \neq 0 $, regardless of $ N_c \pmod 4$, the $ q=4 $ SYK is in GUE.
   The hybrid complex fermion SYK stays in the GUE across $ K/J =1 $ until to $ K/J \sim e^{1.5} $,
   then there is a CIT from GUE to the Poisson as $ K/J $ increases further.

    (c) For $ N_c \pmod 4=2 $, at the half filling $ q_c=0 $,  the complex $ q=4 $ fermion SYK is in GSE at $ K=0 $.
        Because $ P^2=-1 $, any energy level is doubly degenerate, so when doing ELS,
        we only pick up one of the  doubly degenerate levels to demonstrate the GSE at $ K/J=0 $.
        Any small $ K $ breaks the degeneracy, then we may need to consider both sets of energy levels, then
        it is easy to see that any small $ K $ makes $ \langle \tilde{r} \rangle $ small, so $ \langle \tilde{r} \rangle $
        starts from zero and increases as $ K/J $ increases.
        The hybrid complex fermion SYK is in the GUE in a wide range near $ K/J =1 $.
        There is a CIT from GUE to the Poisson as $ K/J $ increases further.

        So in this case, using the NN ELS is not enough.
	One must to use the combination of NN and the NNN ELS presented in Sec.2.
        The $ \langle\tilde{r}'\rangle $ was also shown in Fig.\ref{ComplexSYK_Type1}c.
	It is complete to combine both $ \langle\tilde{r}\rangle $ and $ \langle\tilde{r}'\rangle $ into the same figure.
        When $\langle \tilde{r}\rangle$ is close to be zero,
        the NNN $ \langle\tilde{r}'\rangle $ still shows GSE until $ K/J \sim e^{-5} $.
        When $\langle \tilde{r}\rangle$ reaches the plateau value $ \sim 0.60266 $ of GUE with $ \beta=1 $,
        then according to Eq.\ref{eq:PWW2},
        $ \langle\tilde{r}'\rangle $ reaches an  corresponding plateau value listed in Table I as
        $ \sim 0.7344 $ with a RMT index $ 3 \beta + 1=4 $.
        When $\langle \tilde{r}\rangle$ reaches the plateau value $ \sim 0.38629 $ of the Poisson,
        according to Eq.\ref{eq:PWW2}, $ \langle\tilde{r}'\rangle $ reaches an  corresponding plateau value listed in Table I as
        $ \sim 0.5 $. It is only slightly below the GOE value $ 0.53590 $ with the RMT index $ 3 \beta + 1=1 $.

\section{ Type II hybrid SYK models }

 By Type-II, we mean the integrable side is given by $ (q=2)^2 $ SYK which keeps the PH symmetry Eq.\ref{PH} of the $ q=4 $ SYK.
 We will discuss Majorana and complex type-II hybrid SYK respectively. The results can be contrasted with the corresponding Type-I
 hybrid SYK models.

\subsection{The particle-hole $P$ or $ Z $ conserving hybrid $ q=4 $  Majorana fermion SYK }

  It may also be interesting to study the CIT in the type-II  $ q=4 $ hybrid
  Majorana fermion SYK which keeps the PH symmetry at any ratio of $ K/J $:
\begin{equation}
    H_{M\text{-II},\pm}
	=\sum^{N}_{i<j<k<l} J_{ijkl} \chi_{i} \chi_{j} \chi_k \chi_l
	    \pm [ i \sum^N_{i<j}K_{ij} \chi_{i} \chi_{j} ]^2
\label{mix12}
\end{equation}
where  $J_{ijkl}$, $K_{ij}$ are real and  satisfy the Gaussian distributions with
$\langle J_{ijkl}\rangle=0$, $\langle J^2_{ijkl} \rangle= 3! J^2/N^3 $ and
$ \langle K_{ij} \rangle=0$, $\langle K^2_{ij} \rangle= K/N $ respectively.
Note that here we use $\langle K^2_{ij} \rangle\sim K$  to make $ K /J$ dimensionless.

Obviously, the second term of $H_{M\text{-II},\pm}$ can be written as $ H^2_{M2} $.
In contrast to Eq.\ref{mix1},  it still keeps the $ P $ symmetry.
So symmetry analysis alone can not distinguish between $ H_{M4} $ and $ H^2_{M2} $
despite the former is chaotic, the latter is integrable.
Because $ H_{M2} $ is integrable, so is $ H^2_{M2} $.
This can be most conveniently seen from the NN ratio
$ r_n= s_n/s_{n+1} $ of the $ H_{M2} $.
Then $ R_n $ of the $ H^2_{M2} $ can be written as:
\begin{equation}
  R_n= \frac{ E^2_n - E^2_{n-1} }{ E^2_{n+1} - E^2_{n} }
	\sim \frac{ E_n - E_{n-1} }{ E_{n+1} - E_{n} }= r_n
\label{Rrn}
\end{equation}
where, similar to the cancelation of the density of states in $ r_n $, the center of two NN energies just cancels in the ratio.
Very similarly, one can show that the ratio of the NNN energy spacing
$ R^{\prime}_n \sim r^{\prime}_n $.
So the  ELS of $ H^2_{M2} $ remains Poisson.


  This model with both $ \pm $ sign was studied before in \cite{MBLSPT} by $ 1/N $ expansion  and
  by the RG analysis at $ N=\infty $. By performing the RG on the
  coefficient of the $ H^2_{M2} $ term around
  the $ q=4 $ SYK conformally invariant fixed point, it was found that the $ + $ sign is marginally irrelevant (Fig.\ref{sykrg}b2),
  so the $ q=4 $ non-integrable SYK NFL fixed point is stable in the IR against the $ + H^2_{M2} $ perturbation.
  However,  the $- $  is marginally relevant (Fig.\ref{sykrg}b1),
  it flows to the integrable FL fixed point controlled by $ -H^2_{M2} $.
  However, our ED results show that there is very little differences between the two signs in ELS.
  This fact could be explained as follows:
\begin{align}
-H_{M\text{-II},-}[J, K]
	&=\!\!\!\sum^{N}_{i<j<k<l}\!\! J^{\prime}_{ijkl} \chi_{i} \chi_{j} \chi_k \chi_l
	+ [ i \sum^N_{i<j} K_{ij} \chi_{i} \chi_{j} ]^2 \nonumber\\
	&= H_{M\text{-II},+}[J^{\prime}, K]
\label{pmeq}
\end{align}
   where $   J^{\prime}_{ijkl}=- J_{ijkl} $.
   So $ J^{\prime} $ and $ J $ satisfy the same distribution.

\begin{figure}[!htb]
\includegraphics[width=\linewidth]{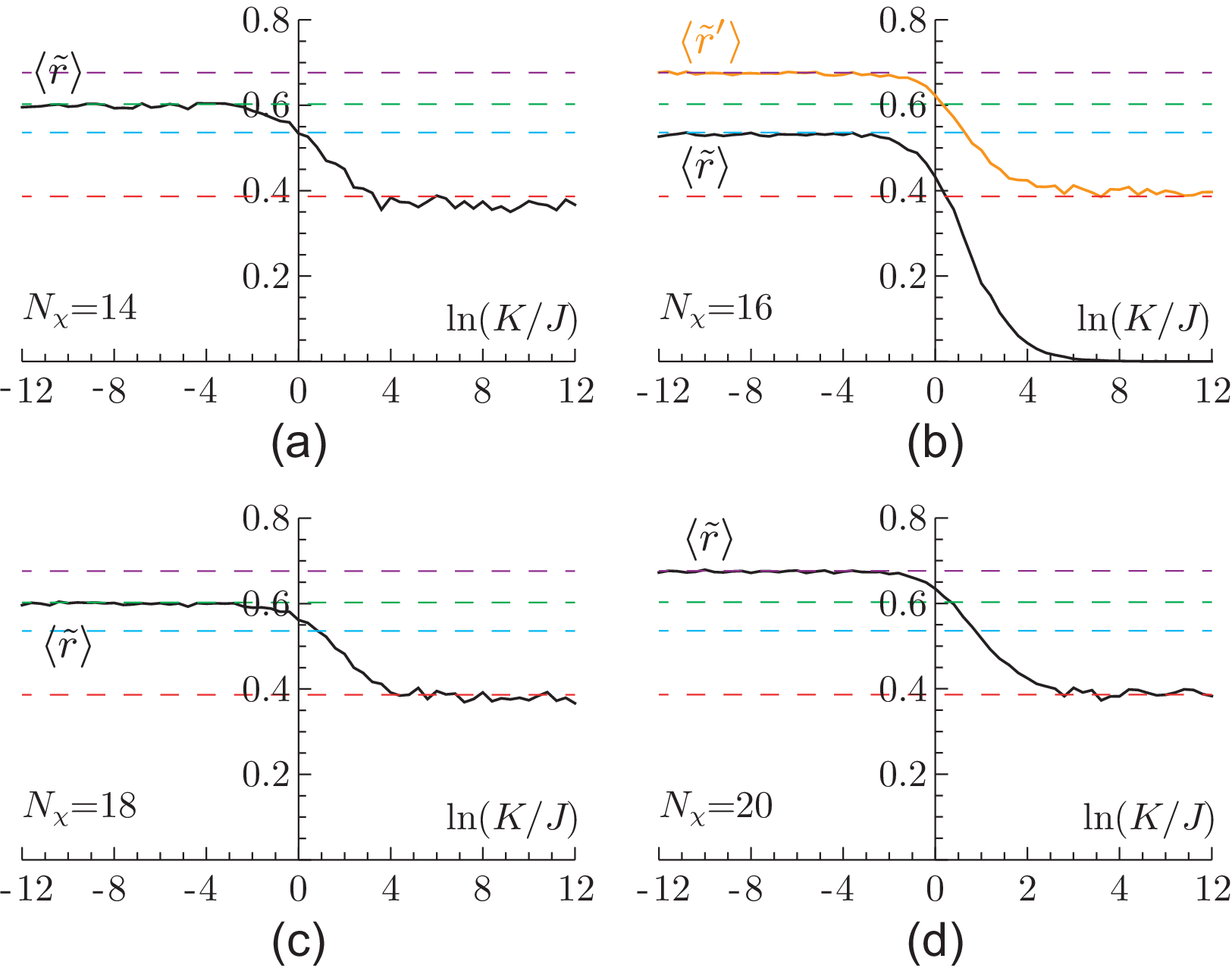}
\caption{ Even $ N_{\chi} $: the evolution of the ELS for type-II hybrid Majorana SYK model when $ N_{\chi} $ is even.
$\langle\tilde{r}\rangle$ (black curve),and $\langle\tilde{r}'\rangle$ (orange curve) in (b)
evaluated for $N_\chi=14,16,18,20$ and are averaged over  $100$, $80$, $60$, $40$  samplings respectively.
Notably, for $ N \pmod 8=0 $ case in (b), there is a double degeneracy at the $ (q=2)^2 $ side,
the NN ratio $ \langle \tilde{r}  \rangle $ is rapidly changing near the $ (q=2)^2 $ side,
so it is quite difficult to determine the KAM regime \cite{KAM} ( See Sec V-C ). Fortunately,
the NNN ratio $ \langle \tilde{r}'  \rangle $ shows a nice plateau regime near the  $ (q=2)^2 $ side, the KAM regime can be
easily identified and listed in the Table III-VI. This dramatic advantage of the new NNN ratio over the known NN ratio were
further demonstrated in all the relevant figures.  For $ N \pmod 8=4 $ case in (d), all the energy levels are doubly
degenerate  at any $ J/K $. The crucial difference between $ \langle \tilde{r}'  \rangle $ in (b) and $ \langle \tilde{r}  \rangle $  in (d)
are stressed in the text.}
\label{MajoranaSYK_Type2_Even}
\end{figure}

\begin{figure}[!htb]
\includegraphics[width=\linewidth]{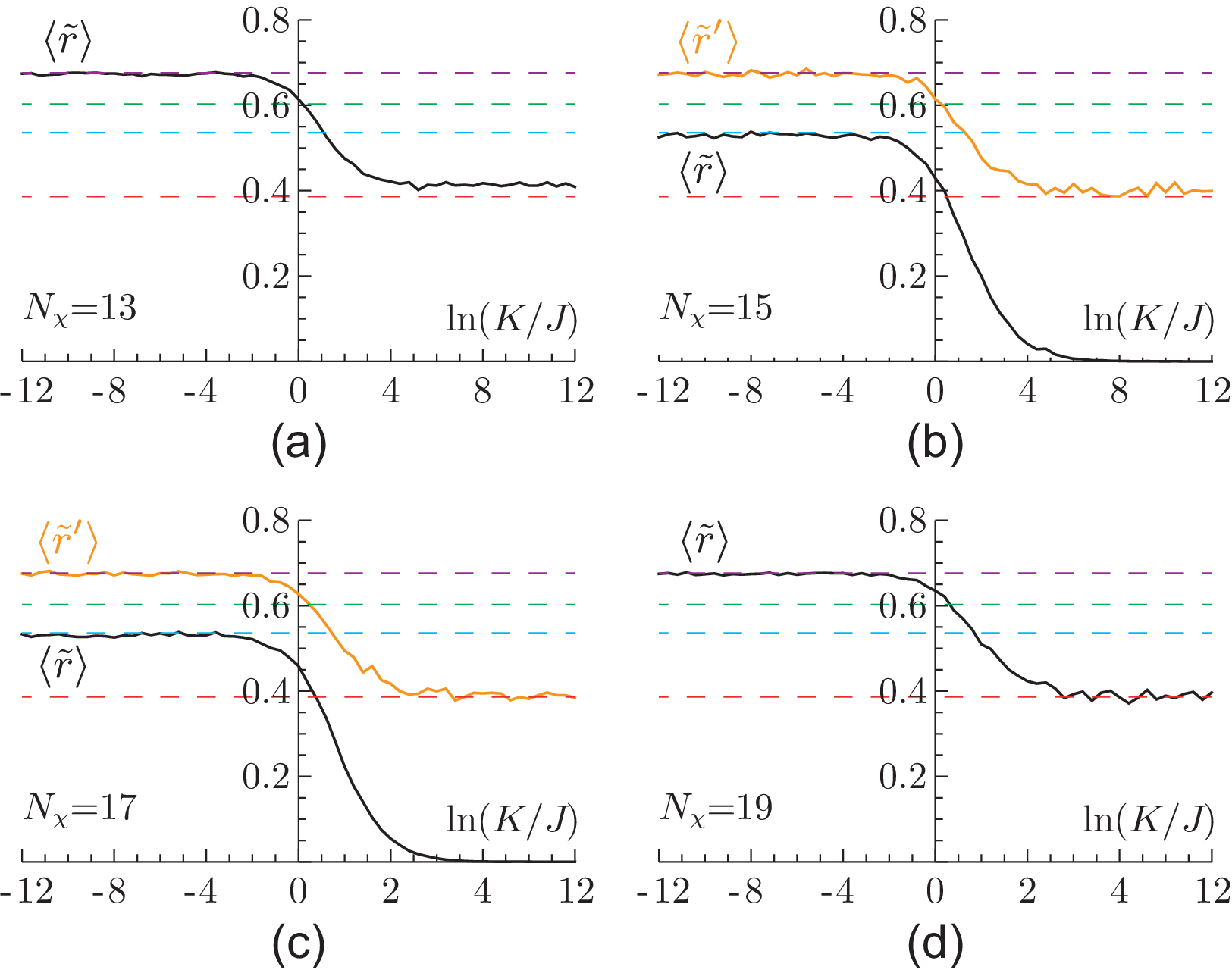}
\caption{ The evolution of the ELS for type-II hybrid Majorana SYK Model when $ N_{\chi} $ is odd.
$\langle\tilde{r}\rangle$ (black curve),
	and $\langle\tilde{r}'\rangle$ (orange curve in (b) and (c) )
	evaluated for $N_\chi=13,15,17,19$ and are averaged over  $100$, $80$, $60$, $40$  samplings respectively.
For $ N \pmod 8=1, 7 $ GOE case in (b) and (c), one can see the advantages of using $ \langle \tilde{r^{\prime}}  \rangle $
over $ \langle \tilde{r}  \rangle $,  especially in the KAM side. For $ N \pmod 8=3, 5 $ GSE case in (a) and (d),  all the energy levels
are doubly degenerate at any $ J/K $.	}
\label{MajoranaSYK_Type2_Odd}
\end{figure}

Let $E^{+}_n$ be an ordered set of energy levels of $ H_{M\text{-II},+}[J, K] $, then
 $s^{+}_n=E^{+}_{n+1}-E^{+}_{n} > 0 $ are the NN spacings.
Then $ E^{-}_n=- E^{+}_n $ is the corresponding ordered set of energy levels
of $ H_{M\text{-II},-}[J^{\prime}, K] $
and $ s^{-}_n=E^{-}_{n}-E^{-}_{n+1}=E^{+}_{n+1}-E^{+}_{n} = s^{+}_n $ are the corresponding NN spacings.
Similarly, the NNN spacing
$ s^{\prime -}_n=E^{-}_{2n-1}-E^{-}_{2n+1}=E^{+}_{2n+1}-E^{+}_{2n-1} = s^{\prime +}_n $.
Because $ ( J^{\prime}, K ) $ and $ ( J, K ) $  satisfy the same distribution,
so we conclude $ H_{M\text{-II}, \pm} $ satisfy the same ELS. \cite{AFM}.
This is confirmed by our ED.
So we just show our $+$ sign results in Fig.\ref{MajoranaSYK_Type2_Even} and \ref{MajoranaSYK_Type2_Odd}.
However, there is an exchange between the ground state and the highest energy state in the $ \pm $ sign,
so the $ H_{M\text{-II}, \pm} $ will have completely different ground states.
This can also be seen by the RG analysis  in Fig.\ref{sykrg}b1,b2  \cite{MBLSPT}.
   This fact may show that the CIT characterized by the RMT at a finite $ N $
   may be complementary to the QPT characterized by the RG at $ N=\infty $.
   Further elucidations on the intricate relation between RG and RMT will be given in Sec. V.

  Because $ [P, H_{M\text{-II},\pm} ]=0 $ or $  [Z, H_{M\text{-II},\pm} ]=0 $ ( when $ N \pmod 8=1, 5 $, one use $ Z $, in all the other cases, one use $ P $ ), so symmetry  analysis alone can not distinguish between $ H_{M4} $ and $ H^2_{M2} $.
  So the symmetry classification in Sec.3 still applies to this Type-II hybrid  Majorana fermion SYK at any $ J/K $.
  So the Table II still holds.  Our ED studies \cite{JW}
  in Fig.\ref{MajoranaSYK_Type2_Even} and \ref{MajoranaSYK_Type2_Odd} shows
  that this is true up to $ K/J \sim e^{-2} $.
  Then the KAM theorem applies at small $ J/K \ll 1 $ where the ELS becomes Poissonian.
  There is a CIT from the corresponding WD to the Poisson as $ K/J $ increases.
  However, the degeneracy in Table I remains true at any $ J/K $.
  In the following, we discuss $ N $ even and odd case respectively.

  For $ N $ even shown in Fig.\ref{MajoranaSYK_Type2_Even},  there is always a CIT from the corresponding WD to the Poisson.
  Several salient features need to be stressed.
  For $ N \pmod 8=0 $ in (b),  one may also look at the ELS from the $ H^2_{M2} $ side when $ J/K \ll 1 $.
  Because $ \{ P, H_{M2} \} =0 $, then $  \psi, P \psi $ are  still in the same parity sector,
  but have two different eigenvalues $ \pm \lambda $, so are orthogonal.
  But $ [ P, H^2_{M2} ] =0 $,  a pair of orthogonal eigenstates $ ( \psi, P \psi) $ have the same eigenvalue $ \lambda^2 $.
  So $ H^2_{M2} $ is doubly degenerate at $ J=0 $.
  However, the double degeneracy is broken by any $ J >0 $.
  Then using the NN ELS is not enough. One must use the combination of NN and the NNN ELS presented in Sec.2.
  In contrast to all the previous cases with $ P $ or $ Z $ violating Type-I hybrid SYK models
  where the  doubly degeneracy comes from the $ q=4 $ non-integrable side,
  here the doubly degeneracy comes from the integrable side.
   It is best to combine $\langle \tilde{r}\rangle$ and $ \langle\tilde{r}'\rangle $
   in Fig.\ref{MajoranaSYK_Type2_Even}b and {\sl read them from the $ ( q=2 )^2 $ integrable side}.
        When $\langle \tilde{r}\rangle$ is close to be zero,
        the NNN $ \langle\tilde{r}'\rangle $ still shows Poisson until $ J/K \sim e^{-4} $.
        When $\langle \tilde{r}\rangle$ reaches the plateau value $ \sim 0.53590 $ of the GOE with $ \beta=1 $,
        then according to Eq.\ref{eq:PWW2},
        $ \langle\tilde{r}'\rangle $ reaches an  corresponding plateau value listed in Table I as
        $ \sim 0.6769 $ which is quite close to GSE with a RMT index $ 3 \beta + 1=4 $.

        For $ N \pmod 8=4 $ in (d),  there is a CIT from GSE to Poisson as $ K/J $ increases.
        As shown in Table II, there is always a double degeneracy at any $ J/K $.
        Although $\langle\tilde{r}'\rangle$ in (b) is quite similar
        to $\langle\tilde{r}\rangle$ in (d), both of which looks like to show a CIT from GSE to Poisson,
        they have very different physical meanings.
        As said above,  $ \langle\tilde{r}'\rangle $ with $ N \pmod 8=0 $ in (b) represents NNN ELS, so it stands  for a CIT from GOE to Poisson.  While  $\langle \tilde{r}\rangle$ with $ N \pmod 8=4 $ in (d) represents NN ELS, so it is  a true CIT from GSE to Poisson.

       For $ N \pmod 8=2,6 $ in (c) and (a),  there is no degeneracy at any $ J/K $. There is a CIT from GUE to Poisson as $ K/J $ increases.

       The odd $ N $ case is shown in Fig.\ref{MajoranaSYK_Type2_Odd}.  As shown in Table I, one
       continue to use $ P $ symmetry when $ N \pmod 8=3, 7 $, but  must use $ Z $ symmetry when $ N \pmod 8=1,5 $.
       Due to the absence of GUE,  there are only two cases: the CIT from GSE to Poisson in (a) and (d)
       and the CIT from GOE to Poisson in (b) and (c).
       They show similar evolution patterns as the corresponding CITs for even $ N $ case shown in Fig.\ref{MajoranaSYK_Type2_Even}.
       Again, although $\langle\tilde{r}'\rangle$ in (b) and (c) are quite similar
       to $\langle\tilde{r}\rangle$ in (a) and (d), both of which looks like to show a CIT from GSE to Poisson,
       they have very different physical meanings. The former is actually a CIT from GOE to Poisson.
       While the latter is a true a CIT from GSE to Poisson with a double degeneracy.

\subsection{The particle-hole  $P$  conserving  hybrid $ q=4 $  complex fermion SYK}

Now we study the CIT in the following $ q=4 $ type-II complex fermion SYK \cite{can}
which also keeps the $ P $ symmetry at any ratio of $ K/J $:
\begin{equation}
H_{C\text{-II},\pm}
	=  \sum^{N}_{i<j,k<l} J_{ij;kl} c^{\dagger}_{i} c^{\dagger}_{j} c_k c_l \pm  [ \sum^N_{i<j}
K_{ij} c^{\dagger}_{i} c_{j} ]^2- \mu q_c
\label{mix22}
\end{equation}
where $ K^{*}_{ij} =K_{ji} $ is  a Hermitian matrix satisfying
$ \langle K_{ij} \rangle=0,  \langle |K_{ij}|^2 \rangle=  K/N $.
Note that here we use $\langle K^2_{ij} \rangle\sim K$  to make $K/J$ dimensionless.

\begin{figure}[!htb]
\centering
\includegraphics[width=\linewidth]{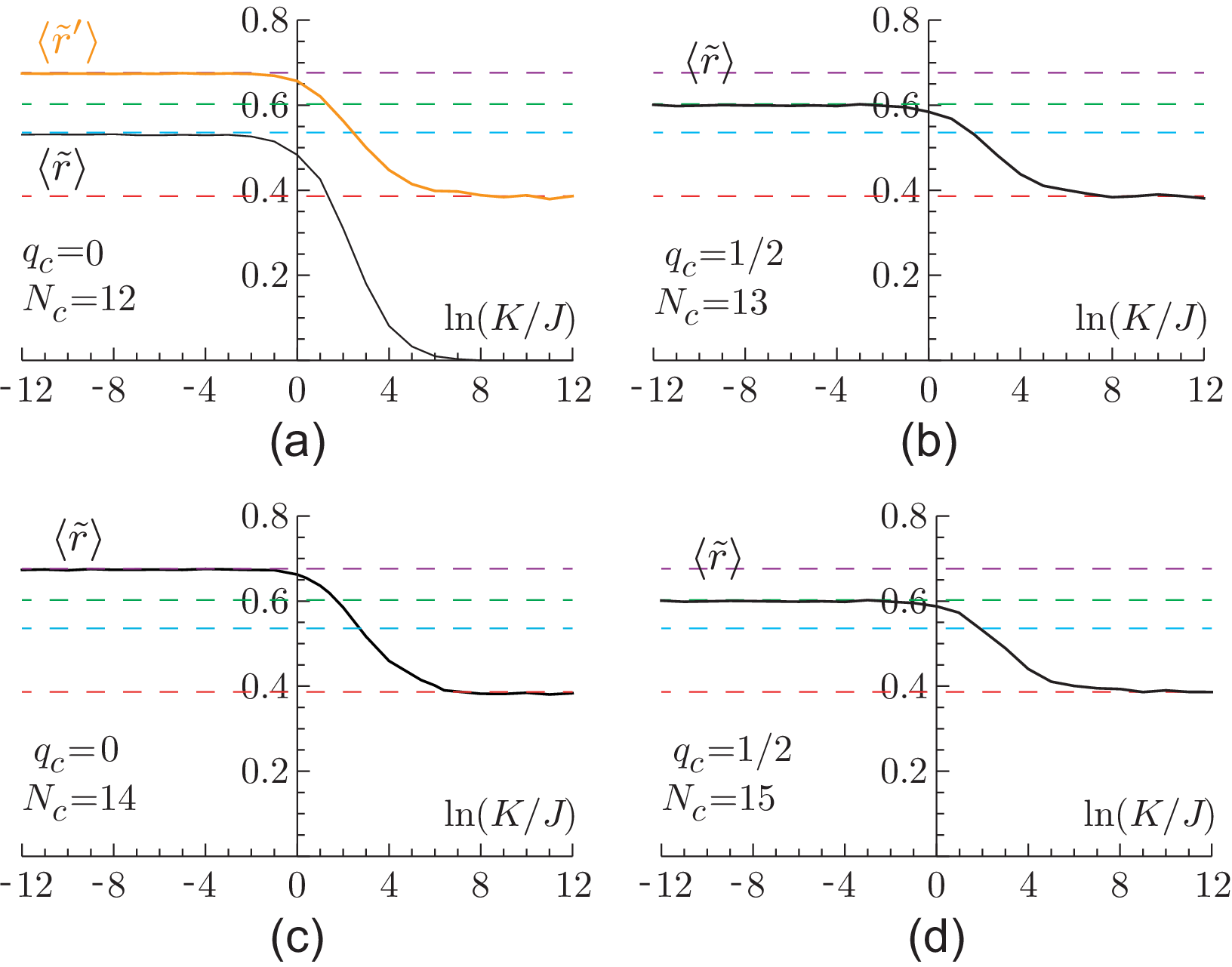}
\caption{  The evolution of the ELS for type-II hybrid complex SYK model.
$\langle\tilde{r}\rangle$ (black curve) and $\langle\tilde{r}'\rangle$ (orange curve) in (a)
	evaluated with $N_c=12,13,14,15$ and are averaged over
	$100$, $50$, $50$, $50$ samplings respectively.	
For $ N_c \pmod 4=0, q_c=0 $ GOE case in (a), one can see the advantages of using $ \langle \tilde{r^{\prime}}  \rangle $
over $ \langle \tilde{r}  \rangle $, especially in the KAM side. For $ N_c \pmod 4=2, q_c=0  $ GSE case in (c), all the energy levels
are doubly degenerate at any $ J/K $.}
\label{ComplexSYK_Type2}
\end{figure}

  Obviously, the second term can be written as $ H^2_{C2} $. So it still keeps the $ P $ and $ Z $ symmetry.
  So symmetry analysis along can not distinguish between $ H_{C4} $ and $ H^2_{C2} $.
  Then the symmetry classification on complex SYK still applies to this $ P $ conserving hybrid $ q=4 $  complex fermion SYK.
  When $ N \pmod 4=0,2 $ and $ q_c=0 $, the ELS is GOE and GSE with the degeneracy $ d=1,2 $ respectively.
  But when $ N \pmod 4=1,3 $ and $ q_c=\pm 1/2 $, it is GUE with the degeneracy $ d=1 $.
  (In fact, as long as $ q_c \neq 0 $,  there is no  $ P $ symmetry anymore,
  so it is in GUE regardless of the value of $ N \pmod 4$).
  Our ED studies \cite{JW}
  in Fig.\ref{ComplexSYK_Type2} shows that this is true until $ K/J \sim e^{-1} $.
  Then the KAM theorem applies al small $ J/K \ll 1 $ where the ELS becomes Poisson.
  However, the degeneracy remains true at any $ J/K $. There is a CIT from the corresponding WD to the Poisson as $ K/J $ increases.

  The rest of discussions are quite similar to the $ P $ or $ Z $ conserving Type-II hybrid Majorana fermion SYK discussed in the last subsection.
  Similar to Eq.\ref{Rrn},  it is easy to show that because $ H_{C2} $ is integrable, so is $ H^2_{C2} $,
  so its ELS remains Poisson.
  Furthermore, the eigenvalue of $ H^2_{C2} $ is always positive.
  Both $ \pm $ sign in the second term need to be considered.
  In the RG sense, we expect the $ + $  ($-$) sign in Eq.\ref{mix22} is irrelevant (relevant).
  However, Eq.\ref{pmeq} adopted to Eq.\ref{mix22} shows the ELS stay the same which is confirmed by our ED.
  So we just show our $ + $ sign results in Fig.\ref{ComplexSYK_Type2}.
  There is always a CIT from the corresponding WD to the Poisson as $ K/J $ increases.

  Notably, for $ N_c \pmod 4=0, q_c=0 $ in (a),  one may also look at the ELS from the $ H^2_{C2} $ side when $ J/K \ll 1 $.
  Because $ \{ P, H_{C2} \} =0 $, but $ [ P, H^2_{C2} ] =0 $, so $ H^2_{C2} $ is doubly degenerate at $ J=0 $.
  Then using the NN ELS is not enough. One may need to use the combination of NN and the NNN ELS presented in Sec.2.
  In contrast to the previous  $ P $ violating type-I hybrid complex SYK model
  where the  doubly degeneracy comes from the $ q=4 $ chaotic side, here
  the doubly degeneracy comes from the integrable side. It is complete to combine
      both $\langle \tilde{r}\rangle$ and $ \langle\tilde{r}'\rangle $ in
      Fig.a and {\sl read them from the $ q=2 $ integrable side}.
        When $\langle \tilde{r}\rangle$ is close to be zero,
        the NNN $ \langle\tilde{r}'\rangle $ shows Poisson until $ J/K \sim e^{-6} $.
        When $ \langle \tilde{r}\rangle $ reaches the plateau value $ \sim 0.53590 $ for the GOE with $ \beta=1 $,
        then according to Eq.\ref{eq:PWW2},
        $ \langle\tilde{r}'\rangle $ reaches an  corresponding plateau value listed in Table I as
        $ \sim 0.6769 $ which is quite close to the GSE with a RMT index $ 3 \beta + 1=4 $.
        There is always a double degeneracy at any $ J/K $ in (c).
       Although $ \langle\tilde{r}'\rangle $ in (a) is quite similar to
       $ \langle\tilde{r}\rangle $ in (c), both of which looks like to show a CIT from GSE to Poisson, they have very different physical meanings.  As said above, $ \langle\tilde{r}'\rangle $ in (a) when $ N_c \pmod 4=0, q_c=0 $ represents NNN ELS, so it stands
       for a CIT from GOE to  Poisson.
       While $\langle \tilde{r}\rangle$ in (c) when $ N_c \pmod 4=2, q_c=0 $ represents NN ELS, so it stands
       for a CIT from GSE to  Poisson and  each energy level is doubly degenerate at any $ K/J $.

\begin{figure}
\centering
    \includegraphics[width=0.65\linewidth]{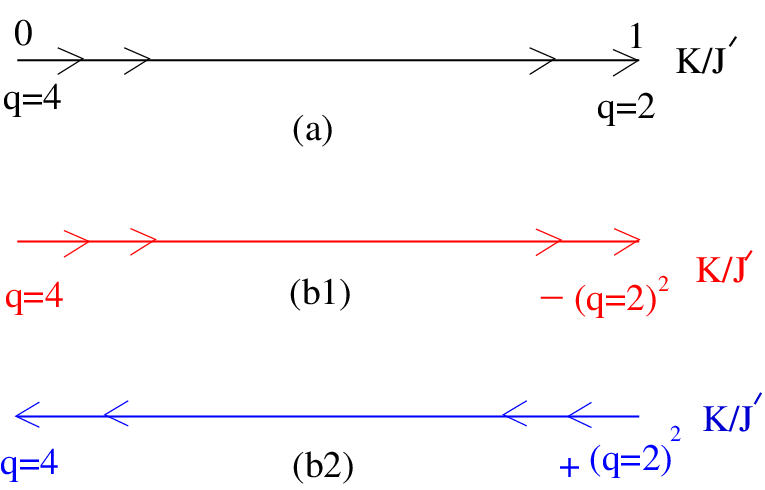}
\caption{RG flow of type-I and type-II hybrid SYK models between two different CFT fixed points.
	(a) The RG flow of the type-I hybrid SYK models.
	The $ K/J^{\prime}, J^{\prime}= \sqrt{ J^2+K^2} $ is relevant to the $ q=4 $ SYK fixed point.
	So the ground state is always a Fermi liquid (FL) with well
	defined low energy quasi-particle excitations.
	(b) Upper, The RG flow of the type-II hybrid SYK models with $-$ sign.
	The $ K/J^{\prime} $ is marginally relevant to the $ q=4 $ SYK fixed point.
	So the ground state is always a Fermi liquid (FL) with well
	defined low energy  quasi-particle excitations.
	Lower, The RG flow of the type-II hybrid SYK models with $+$ sign.
	The $ K/J^{\prime} $ is marginally irrelevant to the $ q=4 $ SYK fixed point.
	So the ground state is always a Non-Fermi liquid (NFL) without the low energy quasi-particle excitations \cite{ctheorem}. }
\label{sykrg}
\end{figure}

\section{Contrast KAM theorem of integrability with stability of quantum chaos  }

     In classical chaos, the classical  Kolmogorov-Arnold-Moser (KAM) theorem \cite{KAM} states that if an integrable Hamiltonian $ H_0 $ is disturbed by a small perturbation $ \Delta H $,
    which makes the total Hamiltonian $ H=H_0+ \Delta H $ , non-integrable. If the two conditions are satisfied:
    (a) $  \Delta H $ is sufficiently small (b) the frequencies $ \omega_i $ of $ H_0 $ are in-commensurate, then the system remains quasi-integrable.   It plays important roles to study the stability of solar systems.
    It remains an outstanding problem to find an quantum analogue of the KAM theorem.
    Here, we define the quantum analog of the classical KAM theorem as the range of a chaotic perturbation which still keeps
    the ELS in the Possionian.
    The quantum KAM theorem in the context of RMT:  For any integrable system subject to an chaotic perturbation,  the KAM theorem holds
    when the chaotic perturbation is below the average many body energy level spacing of the integrable system.

    This theorem should hold to any quantum chaotic systems such as hybrid SYK models or Dicke models \cite{KAM}.
    Instead of providing a rigorous  mathematical proof of this quantum KAM theorem which is left to a future publication \cite{math},
    we will first give some instructive, but naive estimates which lead to an exponential scaling law, then point out it should be replaced by a large power law.  Its precise form may be determined by using
    more rigorous mathematical treatments \cite{GangTian} in a future publication \cite{math}.


\subsection{  Scaling forms of the KAM theorem in the Type-I and Type-II hybrid Majorana or complex SYK }

    For the hybrid Majorana fermion SYK. The size of the Hilbert space is $ 2^{N/2} $.
    On the integrable side, in the $ N \rightarrow \infty $ limit,
    although the single body density of states (DOS) satisfies the semi-circle law,
    the many body DOS satisfies something similar to a Gaussian which has a variance $ \sigma= \sqrt{N/8} $
    and also an exponentially decay in the two tails ( see appendix B ), so
    average bulk many body energy level spacing is
    $ \sim K \sqrt{N/8} \times 2 \times 3/2 \times 2^{-N/2 } \sim K \sqrt{N}  2^{-N/2 } $ where $2/3 $
    takes care of only $ 2/3 $ states are within the width $ \pm \sigma = \sqrt{N/8} $.
    Note that the low or high lying spacing is $ \sim 1/N $  ( which is much larger than the bulk energy level spacing )
    indicating the existence of the quasi-particles.
    When doing the ELS, one can simply throw away these low and high energy levels.
    Because all these many body energy levels are un-correlated and satisfy the Possion distribution,
    so naively we expect that when the chaotic perturbation $ J $ is smaller than this bulk average spacing,
    the KAM theorem holds:
\begin{equation}
     (J/K)_{M} \sim  \sqrt{N} 2^{-N/2} = \sqrt{N} e^{-N\ln2/2}
\label{kamM}
\end{equation}
    This scaling seems matche well with the ED results listed in Table III for Type I and Table V for Type II
    hybrid Majorana SYK. Note that the prefactor $ \sqrt{N} $ may not be important in the large $ N $ limit,
    but is important when comparing with the ED data at the sizes we can perform the ED.
    For example, taking $  N= N_{\chi}=14 $,
    we can see $ J/K \sim 4 \times 2^{-7} \sim 2^{-5}\sim 0.03 $ which is quite close to that listed in the Table III.
    The KAM  regime in  the other sizes seem also fit the scaling law well.
    However, as argued below Eq.\ref{kamC}, the naive Eq.\ref{kamM} should be replaced by a large
    power law $ \sim 1/N^{\alpha_M}, \alpha_M \gg 1 $.
    Unfortunately, the ED at the available sizes may not be able to distinguish the two quite different scaling forms.

    For the hybrid complex fermion SYK. Due to the conserved total charge to be $ N/2 $ at the half filling
    ( other filling can be similar discussed ). The size of the Hilbert space is $  C^{N/2}_N = \frac{ N ! }{ ( \frac{N}{2} !)^2 } $.
    By using Sterling formula $ \ln N != N \ln N-N $ which holds for a large $ N $,
    one can still show $  C^{N/2}_N \sim 2^{N}/\sqrt{N} $ for a large $ N $.
    On the integrable side, in the $ N \rightarrow \infty $ limit,
    although the single body density of states (DOS) satisfies the semi-circle law,
    the many body DOS satisfies something similar to a Gaussian which has a variance $ \sigma= \sqrt{N/4} $
    and also an exponentially decay in the two tails ( see appendix B ), so
    average bulk many body energy level spacing is $ \sim K \sqrt{N/4} \times 2 \times 3/2 \times 2^{-N } \sqrt{N} \sim 3/2 K N 2^{-N } $.
    Again, because all these many body energy levels are un-correlated and satisfy the Possion distribution.
    so naively we expect that when the chaotic perturbation $ J $ is smaller than this average spacing,
    the KAM theorem holds:
 \begin{equation}
     (J/K)_{C} \sim  3/2 N 2^{-N } = 3/2 N e^{-N\ln2 }
\label{kamC}
\end{equation}
    In practice, when comparing the data in  Table IV for Type I and Table VI for Type II,
    it is more accurate to use the actual $  C^{N/2}_N $ instead of its  Sterling form for the sizes we can perform the ED.
    For example, taking $ N=N_c=12 $, then $ C^{6}_{12} \sim 10^{-3} $,
    we can see $ J/K \sim 6 \times 10^{-3} $ which seems close to that listed in the Table IV and VI.
    The KAM  regime in  the other sizes in Table IV and VI seem also fit the scaling law well.

    However, as already alerted below Eq.\ref{kamM}, the naive Eq.\ref{kamC} should be replaced by a large
    power law $ \sim 1/N^{\alpha_C}, \alpha_C \gg 1 $.
    Unfortunately, the ED at the available sizes may not be able to distinguish the two quite different scaling forms.
    The arguments go as follow: in the integrable side,  taking two nearest neighbouring (NN) bulk states $ | B1 \rangle= |n_1,n_2,\cdots,n_N\rangle_{B1} $
    with the eigen-energy $ E_{B1}= \sum_i n^{B1}_i \epsilon_i $ and $
    |B2 \rangle = |n_1,n_2,\cdots,n_N\rangle_{B2} $  with the eigen-energy $ E_{B2}= \sum_i n^{B2}_i \epsilon_i $.
The NN means the eigen-energy difference $ E_{B1}- E_{B2} $ is of the many-body origin  Eq.\ref{kamC}.
The single particle spacing in the integral side $ H_2 $ is $ \epsilon_2 -\epsilon_1 \sim K/N $.
So $ | B1 \rangle, |B2 \rangle $ must differ by $ \sim N/2 $ particle occupations $ n_i $ to get such a small spacing listed in Eq.\ref{kamC}.
However, every chaotic perturbation $ H_4 $ involves only $ 4 $ particle moves in $ n_i $, so typically one need $ \sim N/2 \times 1/4 = N/8 $ steps
to connect the two NN bulk states. The energy
of intermediate states involve only 4 single particles, so the change of energy could at most be $ 4 K/N $,
then by $ N/8 $ steps of chaotic perturbation which connects $ | B1 \rangle $ to $ |B2 \rangle $, one can estimate the
off-diagonal matrix element $ \langle B1| H_4 | B_2 \rangle  \sim J ( J/(4K/N))^{N/8}  \sim J ( N J/K)^{N/8}  $ which is also
exponentially small when $ NJ/K \ll 1 $.
Due to the PH symmetry at the half filling and the self-average in the quenched disorders in the large $ N $ limit,
one also expect that the diagonal energy shift is tiny
$ \langle B1| H_4 | B_1 \rangle - \langle B2| H_4 | B_2 \rangle\sim 0 $. As argued in \cite{KAM}, the diagonal energy shift 
does not lead to the change of ELS anyway.
The energy level repulsion is solely due to the off-diagonal matrix element $ \langle B1| H_4 | B_2 \rangle  $.
So we expect the naive Eq.\ref{kamC} should be replaced by a large power law $ \sim 1/N^{\alpha_C} $.
The large power $ \alpha_C  \gg 1 $ will be determined by a rigorous mathematical derivation in \cite{math}.
A similar analysis can be applied to the Majorana fermion case to determine the large power  $ \alpha_M $ in Eq.\ref{kamM}.

\subsection{ The lower bound of the dual form of the KAM theorem in the Type-I  hybrid Majorana or complex SYK }

    As mentioned in Sec.III-A, one may also study the quantum analog of the KAM theorem from a dual point of view, namely,
    the stability of quantum chaos of the quantum chaotic $ q=4 $ SYK against the integrable perturbation as one turns on $ K/J $.
    What would be the finite size dependence of the stability of quantum chaos in terms of  $  K/J $ ?
    The many body DOS  of a RMT satisfies the semi-circle law.
    However, as shown in the previous works\cite{Mald,randomM0,randomM},
    the many body DOS of the $ q=4 $ Majorana SYK does not fit the semi-circle law well, but still close to its form near the band edge.
    The $1/N $ expansion can only get the low end of the DOS which scales as $ \sqrt{E} $.
    Its whole form can only be achieved by ED.
    So we use the results in the ED achieved in \cite{Mald,randomM0,randomM}:
    the bandwidth is $ 2 \epsilon^{M}_0 NJ $ where $ \epsilon^{M}_0 \sim 0.05 $ is the ground state energy per site in the unit of $ NJ $.
    Then the average many body energy level spacing is $ \sim 2 \epsilon^{M}_0 NJ  2^{-N/2 } $ which defines the
    Heisenberg time $ t_H \sim \frac{1}{NJ} e^{N} $.
    Note that the low or high lying spacing is $ \sim e^{-s_0 N } $ ( which is still much larger than the bulk energy level spacing )
    indicating the absence of the low energy quasi-particles.
    In sharp contrast to the KAM theorem in the integral side,
    all these many body energy levels are correlated and satisfy various WD distributions,
    so the arguments used to establish the KAM Eq.\ref{kamM},\ref{kamC} on the integral sides break down.
    We expect that when the integrable perturbation $ K $ is smaller than this average spacing,
    it remains the same ELS. Naively, the lower bound of the dual form of the KAM takes
\begin{equation}
     (K/J)_{M}  >  2 \epsilon^{M}_0 N 2^{-N/2} \sim  2 \epsilon^{M}_0 N  e^{-N\ln2/2}
\label{dualM}
\end{equation}

  This lower bound of the scaling seems match well with the ED
  results for GOE and GSE cases listed in Table III for Type I.  For example, taking $ N_{\chi}=16 $,
  we can see $ (K/J)_M > 0.05 \times 32 \times 2^{-8}  \sim 6 \times 10^{-3} $ which is indeed below that listed in the Table III.
  The bound of the dual KAM also seem also fit the other sizes in Table III.


  There is a recent $1/N $ evaluation of the many body DOS of the $ q=4 $ complex SYK at or away from half-filling
  which consists of the contributions from both the re-parametrization mode and the $ U(1) $ charge mode \cite{u1dos}.
  Again, this $1/N $ evaluation only captures the physics near to the band edge $ E_0 $ of the DOS which was shown to be quite similar to
  that in the Majorana fermion case, namely the RMT behaviour $ \sqrt{E-E_0} $.
  However, its whole form at the half-filling can only be obtained by the ED which
  is shown in the appendix C Fig.\ref{complexdos}: it is quite close to be a semi-circle
  with the bandwidth $ 2 \epsilon^{C}_0 NJ $ where $ \epsilon^{C}_0 \sim 0.15 $ is the ground state energy per site in the unit of $ NJ $.
  Then the average many body energy level spacing is $ \sim 2 \epsilon^{C}_0 NJ  \sqrt{N} 2^{-N } $.
  Then Eq.\ref{dualM} can be simply replaced by:
\begin{equation}
     (K/J)_{C}  >  2 \epsilon^{C}_0 N^{3/2} 2^{-N} \sim  2 \epsilon^{C}_0 N^{3/2}  e^{-N ln2}
\label{dualC}
\end{equation}

  This lower bound of the scaling seems match well with the ED
  results for GOE and GSE cases listed in Table IV for Type I.  For example, taking $ N=12 $,
  we can see $ (K/J)_C > 2 \times 0.15 \times 12 \times 10^{-3}  \sim 3.6 \times 10^{-3} $
  which is indeed below that listed in the Table IV.  The dual form seems also hold in the other sizes in the Table IV.

  Due to the strong energy level correlations among the energy levels on the $ q=4 $ chaotic side,
  we expect that the lower bounds Eq.\ref{dualM},\ref{dualC} can be improved significantly.
  In fact, by using the ETH of the two NN bulk states in the chaotic side in Sec.VI-C,
  one can show the diagonal energy shift is tiny shown in Eq.\ref{diagC}.
  Again, as argued in \cite{KAM}, the diagonal energy shift does not lead to the change of ELS anyway.
  The energy level repulsion is solely due to the off-diagonal matrix element
  $ \langle B1| H_2 | B_2 \rangle  $ in Eq.\ref{H12C} which need to be evaluated by some rigorous mathematical treatments \cite{math}.
  A similar analysis can be applied to the dual Majorana fermion case.

  Of course, Eq.\ref{dualM},\ref{dualC} do not apply to the GUE case for Type I  and any case in  Type II. In all these cases,
  there are no changes in ELS from $ q=4 $ side to the bulk, there is a direct CIT from the corresponding WD in the $ q=4 $ side to the
  Possion at the $ q=2 $ side.  The stability of the quantum chaos does not vanish in the thermodynamic limit.

\subsection{ Contrast numerical data with the scaling form of the quantum KAM theorem and its dual form  }

Now from Fig.3-8 we summarize the stability of quantum chaos near $ q=4 $ SYK with the KAM theorem near the integrable $ q=2 $ SYK in type-I and  $ ( q=2 )^2 $ SYK in type-II in the following 4 Tables from which one can conclude the data support
the KAM theorem and the lower bound of its dual form presented in the last subsection.

(1) The validity regime of KAM theorem seem comparable in type-I and type-II.
    In all the cases,  its validity regime  gets smaller in  exponentially fast as the system size gets larger.
    These facts are consistent with the KAM theorem in the hybrid SYK models  conjectured above.

(2) In type-I,  the stability of GOE is comparable to that of GSE. They are also consistent with
    the KAM in the sizes we studied, also approaches zero as the system's size increases.
    This is consist with the lower bound of the dual form of the KAM theorem conjectured above. The GOE or GSE will turn into
    GUE when $ K/J $ increases further.

    However, the stability of the GUE is much more robust than the two. It is also much more robust than the KAM.
    This is due to the fact that GUE is dictated by the symmetry classification anyway, so it
    exists as an intermediate regime in all the three cases, then the stability regime of the GUE at the
    $ q=4 $ side is greatly enhanced. Obviously, it remains a finite value in the thermodynamic limit.


(3) In type-II,  the stability of  quantum chaos is even among GOE, GUE and GSE.
    This is because they are dictated by the symmetry classification at the corresponding $ N $ values anyway.
    They are also much more robust than the KAM theorem.

\begin{table}[!htb]
\caption{
Contrast the stability of quantum chaos at $ q=4 $
with the KAM theorem of integrability at $ q=2 $ for
type-I hybrid Majorana fermion SYK. Notations: $ N_{\chi} $ is the number of Majorana fermions, ELS is Energy Level Statistics, QCS is Quantum Chaos stability. KAM is Kolmogorov-Arnold-Moser theorem. The same notations are used in the following tables.
The KAM shrinks as the size increases. The QCS of the GUE  is greatly enhanced. }
\resizebox{\linewidth}{!}{
\begin{tabular}{ |*{5}{c|} }
\hline\hline
$N_{\chi}$   &14	&16	&18	&20   \\
\hline
ELS at $K=0$ &GUE	&GOE	&GUE	&GSE\\
\hline
QCS: $K/J$   &$e^{0}=1$	&$e^{-5.5}\approx0.004$	    &$e^{0}$	&$e^{-5.5}$ \\
\hline
KAM: $J/K$   &$e^{-3}\approx0.05$ &$e^{-3.5}\approx0.04$ &$e^{-4}\approx0.02$ &$e^{-5}\approx0.01$\\\hline\hline
$N_{\chi}$   &13	&15	&17	&19   \\
\hline
ELS at $K=0$ &GSE	&GOE	&GOE	&GSE\\
\hline
QCS: $K/J$   &$e^{-4.5}\approx0.01$ &$e^{-5}\approx0.007$ &$e^{-5.5}\approx0.004$ &$e^{-5.5}$ \\
\hline
KAM: $J/K$   &$e^{-3}\approx0.05$ &$e^{-3.5}\approx0.04$ &$e^{-4}\approx0.02$ &$e^{-5}\approx0.01$\\
\hline\hline
\end{tabular}
}
\end{table}

\begin{table}[!htb]
\caption{
Contrast the stability of quantum chaos at $ q=4 $
with the KAM theorem of Integrability at $ q=2 $ for
Type-I hybrid complex fermion SYK. The QCS of the GUE  is greatly enhanced. }
\resizebox{\linewidth}{!}{
\begin{tabular}{ |*{5}{c|} }
\hline
$(N_{c},q_c)$	&$(12,0)$		&$(13,1/2)$		&$(14,0)$	&$(13,1/2)$\\  \hline
ELS at $K=0$	&GOE			&GUE			&GSE		&GUE	\\  \hline
QCS: $K/J$	&$e^{-5.5}\approx0.004$	&$e^{1.5}\approx4.5$	&$e^{-6}\approx0.002$	&$e^{2}\approx7.3$\\ \hline
KAM: $J/K$	&$e^{-5}\approx0.008$	&$e^{-5}$		&$e^{-5.5}$	&$e^{-6}$\\ \hline
\end{tabular}
}
\end{table}

\begin{table}[!htb]
\caption{
Contrast the stability of quantum chaos at $ q=4 $
with the KAM theorem of Integrability at $ q=2 $ for
Type-II hybrid Majorana fermion SYK. The QCS remains even among all WD and are much more robust than KAM. }
\resizebox{\linewidth}{!}{
\begin{tabular}{ |*{5}{c|} }
\hline
$N_{\chi}$	&14	&16	&18	&20	\\  \hline
ELS at $K=0$	&GUE	&GOE	&GUE	&GSE	\\  \hline
QCS: $K/J$	&$e^{-2}\approx0.36$ &$e^{-2}$ &$e^{-2}$ &$e^{-2}$\\	\hline
KAM: $J/K$	&$e^{-3}=0.05$	&$e^{-4}=0.018$ &$e^{-5}=0.007$ &$e^{-6}=0.002$	\\
\hline\hline
$N_{\chi}$	&13	&15	&17	&19	\\  \hline
ELS at $K=0$	&GSE	&GOE	&GOE	&GSE	\\	\hline
QCS: $K/J$	&$e^{-2}\approx0.36$ &$e^{-2}$ &$e^{-2}$ &$e^{-2}$\\	\hline
KAM: $J/K$	&$e^{-3}=0.05$	&$e^{-4}=0.018$ &$e^{-5}=0.007$ &$e^{-6}=0.002$	\\
\hline\hline
\end{tabular}
}
\end{table}

\begin{table}[!htb]
\caption{
Contrast the stability of quantum chaos at $ q=4 $
with the KAM theorem of Integrability at $ q=2 $ for
Type-II hybrid complex fermion SYK. The QCS remains even among all WD and are much more robust than KAM. }
\resizebox{\linewidth}{!}{
\begin{tabular}{ |*{5}{c|} }
\hline
$(N_{c},q_c)$	&$12,q_c=0$		&$13,q_c=1/2$		&$14,q_c=0$	&$13,q_c=1/2$\\  \hline
ELS at $K\!=\!0$    &GOE		&GUE			&GSE		&GUE	\\  \hline
QCS: $K/J$	&$e^{-2}\approx0.14$	&$e^{-2}$		&$e^{-2}$	&$e^{-2}$\\  \hline
KAM: $J/K$	&$e^{-6}\approx0.002$	&$e^{-7}\approx0.001$	&$e^{-7}$	&$e^{-8}\approx0.0003$\\  \hline
\end{tabular}
}
\end{table}

  It remains interesting to test the KAM theorem applied to the hybrid
  SYK models and its dual form in much large size systems.
    Unfortunately, the ED can not get to much larger size than $ N=20 $ achieved in this paper.
   In Type-I Majorana fermion hybrid SYK, in the GOE and GSE cases such as Fig.3b,d for even $ N $, Fig.4 for odd $ N $,
   Type-I complex fermion hybrid SYK, Fig.5a,c, we only identified the KAM theorem from the integrable side
   and the low bound of its dual form from the quantum chaos side. However, there is no physical understanding on the
   two ends of the intermediate GUE plateau.
  The two crossover regimes near the two ends may not satisfy ETH.
  As alerted at the end of Sec.V-B, in the GUE case ( on the $ q=4 $ SYK side ) of the type-I SYK and
  all the type-II hybrid SYK models, there is no change from the ELS from the $ q=4 $ SYK side to that in the bulk of the
  hybrid SYK models, so the quantum chaos is always much more robust against an
  integrable perturbation than the other cases.
  There is no analytic estimate on the stability regime
  of this robust quantum chaos. Of course, in contrast to the KAM and its dual form, this robust quantum chaos regime stays robust
  in the thermodynamic limit.

\section{ Bulk ( or intermediate ) states characterized by the RMT versus Edge ( namely, low and high energy )
states captured by  RG or $1/N $ expansion }

It was well established that in terms of symmetries and the space dimension,
the RG (including the DMRG, MPS and tensor network)
can be used to classify many body quantum phases and quantum phase transitions at $ T=0 $
and classical phase transitions at finite temperatures \cite{aue,subirbook}.
The RG focus on infra-red (IR) behaviours of the system
which are determined by the ground state and low energy excitations.
The RG is also intimately connected to general relativity through the holographic principle \cite{holoRG}.
RG is usually applied to a system in the thermodynamic limit to characterize the  quantum phase transitions (QPT).
When it gets to a finite system, finite size scalings near the QCP can be used to extract the information at a finite $ N $
from the knowledge at $ N=\infty $. For a zero dimensional system such as  SYK models or $ U(1)/Z_2 $ Dicke models \cite{KAM}
with infinite-range interaction,
one can use a $ 1/N $ expansion to study the similar properties.

While the 10-fold way RMT classification scheme to describe quantum chaos in a many body system
only depends on two anti-unitary  or one unitary chiral operators, seems independent of space dimension.
It covers all the energy levels of the system, so can be used to characterize the CIT.
RMT is usually applied to a system with a finite size $ N $.
In a completely different context, the 10-fold way can also be used to
classify topological equivalent class for
non-interacting ( single particle ) topological insulators and topological superconductors \cite{kane,zhang,10-fold4}.
In this case, it also depends on the space dimension and has the Bott periodicity $d\to d+8$.

\subsection{  QPT versus CIT }

The differences and connections between RG at $ N =\infty $, then a $ 1/N $ expansion at a finite $ N $ and the RMT at a finite $ N $
are explicitly demonstrated in the two types of SYK models in this work.
For example, in the thermodynamic limit $N=\infty$,
the fermions in the type-I hybrid SYK models have
scaling dimension $1/2$ and $1/4$ at the $q=2$ and $q=4$ SYK respectively.
So the $q=2$ ($q=4$) SYK is a stable (unstable) conformably invariant fixed point
(Fig.\ref{sykrg}a).
Any $ J $ ($ K $) is irrelevant (relevant) to the $ q=2 $ ($q=4$) SYK.
At any $ K/J \neq 0 $, the ground state is controlled by the $ q=2 $ SYK fixed  point,
so it is always a non-chaotic Fermi liquid with well defined low energy quasi-particle excitations.
There is no quantum phase transition (QPT) in the type-I hybrid SYK models.
There is no finite temperature transitions either.
However, as shown in Fig.3-8, there is always a chaotic to integrable transition (CIT) from the GUE to Poisson.
Of course, the name of CIT  need to be interpreted correctly, because it is dramatically different than a QPT,
The former is characterized by RMT, the latter is by RG or $ 1/N $ expansion.
As shown in these figures, if there is a symmetry change from the $ q=4 $ SYK to the hybrid SYK, there are also crossovers between different WD ensembles as $ K/J $ increases.

Quantum or topological phase transitions always start from the zero temperature $ T=0 $,
then raise up to a low temperature $ 1 \ll \beta J < N $. Especially one can find scaling functions for various physical quantities near
the QPT \cite{subirbook}. If putting the system at a finite size, one may also do finite size scaling to identify the QPT
by numerical simulations. It is a bottom/up approach. On the other hand, the
Quantum chaos can only happen at non-zero temperature. In fact, it could even start from the infinite temperature $ T=\infty $, then lower down to the low temperature $ 1 \ll \beta J < N $. So it is a top/down approach.
Therefore the two approaches are complementary to each other.

   QPT only involves the change of the ground states and the low energy excitations,
   while the high energy states are irrelevant. There is a divergent length scale
   ( the time scale is related to the length scale by a dynamic exponent $ z $ )
   and associated scalings near a QPT in the thermodynamic limit $ N =\infty $.
   While the ELS involves the bulk energy levels at a finite but large enough $ N $,
   the low or high energy levels are irrelevant ( see appendix D ).
   So the QPT characterized by RG and the CIT characterized by the RMT are dramatically different, but complimentary to each other.
   Despite the absence of QPT in the two types of the hybrid SYK models presented in Sec.III and IV respectively,
   there could still be a CIT which is dramatically different than  QPT.
   There is no divergent length (or time) scales, no associated scalings, therefore no finite size scalings near a CIT \cite{stupid},
   but there is a KAM in the integrable side and its dual form in the chaotic side if there is a change of level statistics as presented in
   Sec.V.

In the Type-II hybrid SYK models, the coefficient of $ (q=2)^2 $ is marginally irrelevant/relevant when the sign is $+/-$ (Fig.\ref{sykrg}b).
So the ground state with $ + $ sign is the non-Fermi liquid without quasi-particle excitations controlled by $ q=4 $ SYK fixed point,
   while that with $ + $ sign is the Fermi liquid with well defined quasi-particle excitations controlled by $ ( q=2 )^2 $ SYK fixed point \cite{tran2}.
   There is no quantum phase transition (QPT) and no finite temperature transitions either in the type-II hybrid SYK models.
   However, independent of the $ +/- $ sign, there is always a CIT from the corresponding WD of $ q=4 $ SYK to
   the Poisson as shown in Fig.\ref{MajoranaSYK_Type2_Even},\ref{MajoranaSYK_Type2_Odd} and \ref{ComplexSYK_Type2}.
   In fact, as explained in Sec.3.3,3.4, despite the $ +/- $ sign leads to dramatically different ground states, the ELS is identical \cite{AFM}.
   So here we provide an interesting example where the ELS are the same,
   but the ground states are dramatically different \cite{ctheorem}. This fact demonstrates explicitly the dramatic differences
   between the two classification schemes which are complementary to each other, the RG fucus on the ground state and low or high energy excitations, while the RMT focus on the bulk high energy levels ( see appendix D ).

   For another example, as shown in appendix A,  the ground states of the hybrid bosonic SYK is a quantum spin glass which breaks ergodicity.
   There is a finite temperature phase transition at $ T=T_{QSG} $ from the QSG to a paramagnet where the ergodicity is restored.
   However, as shown in Fig.\ref{bcross} in the appendix A, the ELS stays as GOE or GUE. There is no CIT.
   In order to probe the quantum spin glass ground state at  $ T < T_{QSG} $, one need to focus on the low  energy excitations
   which, as alerted above, can not be described by the RMT, need to be investigated by RG or $1/N $ expansion.

In any cases, the RMT of the bulk states are quite insensitive to these edge ( low or high energy ) states\cite{edgediffer}. Namely,
when evaluating the ELS, incorporating or throwing away these edge states will not affect the ELS.
To see these edge states,  RG or $1/N $ expansion may be used to study the ground state and its low or high energy excitations.
In fact, as outlined in the introduction,
using $1/ N $ expansion, many previous works studied the conformably invariant QSL  and also its extensive low energy excitations
with the energy spacing $ \sim e^{-s_0 N } $ which
leads to the zero temperature entropy $ s_0 $ ( taking $ T \rightarrow 0 $ limit after taking $ N \rightarrow \infty $ limit which belongs to
the regime (b) in the following subsection ).

\subsection{ Remarks on Lyapunov exponents from OTOC and SFF in the hybrid SYK models }

   As stressed in the conclusion of Ref.\cite{period},
   there are at least two completely ways to characterize the quantum chaos.
   One way is through evaluating the regulated  out-of-time-ordered correlation function ( OTOC ).
   Another way is to use the RMT to characterize the quantum chaos at a very late time. As demonstrated in all the previous sections,
   when collecting the ELS of the bulk energy levels, low and high energy levels can be simply thrown away
   without affecting the bulk ELS ( see appendix D ). So the two ways are complementary to each other.

   The OTOC  of the two bosonic or fermionic   Hermitian operators $ V^{\dagger}=V, W^{\dagger}=W $ is defined by:
\begin{equation}
  F(t)  =   Tr[y W^{\dagger}(t) y V^{\dagger}(0) y  W(t) y V(0) ]
\label{OTOC}
\end{equation}
  where  $ y=e^{- \beta H/4 }/Z^{1/4} $ is one quarter of the density matrix and
  $ Z(\beta) =Tr e^{-\beta H } $ is the partition function. In the SYK models, $ V=W= \psi $ or $ c $
  for Majorana or complex SYK respectively.
   Its early time exponential growth can be characterized by
   a quantum Lyapunov exponent, while its late time approaches a constant, so become featureless.
   The early time behaviour is mainly determined by the ground state and the low energy excitations,
   while the bulk energy levels are irrelevant ( see appendix D ).
   The OTOC reflects the ground state and low energy excitations, so it is directly related to
   the RG description and can be addressed by $1/N $ expansion.
   It seems quite in-sensitive to the 10-fold way global discrete symmetry classification, therefore
   independent of $ N \pmod 8$  or $ N_c \pmod 4$.

   For the $ q=4 $ Majorana SYK model, as shown in \cite{longtime2}, at any finite temperature,  one must consider the effects of
the finite temperature $ \beta J $ versus the finite size $ N $.
(a) At very low temperatures $ \beta J  > N $, the OTOC takes a power law
$ \sim t^{-6} $ in the long time limit $ t_H > t > \beta > N/J $, so the Lyapunov exponent $ \lambda_L $ can not even be defined at such a low temperature.
(b) At intermediate temperatures $ 1 < \beta J  < N $, then the OTOC takes the exponential form in the early time  upto the Ehrenfest
( or the scrambling time $ t_s \sim \beta \log N $  )
which defines the Lyapunov exponent $ \lambda_L $, but still decays as $ \sim t^{-6} $ in the long time limit  $ t_H > t > N/J > \beta  $.
(c) In the high temperature range $ \beta J < 1 \ll N  $, the physics seems dominated by the
   microscopic energy scale $ J $, the Lyapunov exponent $ \lambda_L \sim J $.

It remains interesting to study how the $ K $ term changes the behaviours of the OTOC and the associated Lyapunov exponent $ \lambda_L $ at
all the three temperature regimes (a),(b) and (c). In both types of hybrid SYK models, the three temperature regimes are still determined by the
competition between the finite size $ N $ and the finite temperatures, so adding a $ K $ term will not change such a division.
(a) is still not the regime to even define a Lyapunov exponent, so we only need to focus on (b) and (c).
In the regime (b) $ 1 \ll \beta J  < N $. We expect that for both type I and type II hybrid SYK models,
the Lyapunov exponent $ \lambda_L > 0 $ can be also computed in all the quantum chaotic regimes
defined in the RMT sense.
We expect $ \lambda_L=0 $ in the KAM regime. In the regime (c), we expect that  in the quantum chaotic side,
any small $ K $ will reduce the Lyapunov exponent to $ \lambda_L \sim J - K + \cdots $.
It will vanish in the KAM regime. Because at such a high temperature, all the energy levels are involved, so we expect
there may exist a one to one correspondence between $ \lambda_L $ and the KAM theorem from the RMT in the regime (c).

   As said in the introduction, in addition to the ELS in the RMT classifications,
   the CIT may also be dynamically diagnosed from  the spectral form factor (SFF) \cite{randomM,wavefunction}:
\begin{equation}
  g(t,\beta)=\frac{ \langle Z(\beta,t)  Z^{*}(\beta,t) \rangle_J }{ \langle Z(\beta) \rangle^2_J }
\label{sff}
\end{equation}
  where $ Z(\beta,t)=Tr( e^{-(\beta+it)H} ) $ and the disorder average was
  taken separately in the enumerator and denominator.

   The  SFF at $ 1 < \beta J < N $ may also be used to measure the dynamic (time-dependent) chaotic behaviours of the
   two types of hybrid SYK models.  A slope-dip-ramp-plateau structure in the time evolution was considered to be evidence
   for the chaotic behaviours. This feature should disappear in the KAM regime.
   It remains interesting to study its evolution in the quantum chaotic regime in both regimes (b) and (c).

   The constraints of the symmetries $ P $ in Eq.\ref{PH} and $ Z $ in Eq.\ref{z} put on the OTOC Eq.\ref{OTOC}  and the
   SFF \ref{sff} will also need to be explored.

\subsection{ Remarks on Eigenstate Thermalization hypothesis (ETH):
its power on bulk states and inability to encode the edge states }

 It is interesting to note that the RMT was originally proposed to study statistically the
 many body energy level correlations of a nuclei with a large atomic number to hold large number of electrons \cite{WD1,WD2}. Then it was used to classify the quantum chaos of non-interacting
 electrons moving in a random potential which may show metal to Anderson insulator transition \cite{efetov}.
 There is a corresponding Chaotic to Integrable transition (CIT) where {\sl the single particle }
 ELS satisfies WD in the metal, while Poisson in the Anderson insulator.

 Recently, there is a renewed interest on Eigenstate Thermalization hypothesis
  (ETH) in many body interacting systems \cite{ETHrev}. It states that
 for any excited ( also called bulk ) state $ |\psi \rangle $ with eigen-energy $ E $ which is above the ground state energy by a finite
 amount in the thermodynamic limit: $ \lim_{V \rightarrow \infty} \frac{E-E_0}{V} \neq 0 $, one may define
 a temperature $ \beta $ corresponding to the state: $ \langle \psi | H | \psi \rangle = Tr H e^{-\beta H }/Z $ where
 $ Z= Tr e^{-\beta H } $ is the partition function,
 then ETH implies than for any local operator $ O $:
\begin{equation}
  \langle \psi | O | \psi \rangle = Tr O e^{-\beta H }/Z
\label{eth}
\end{equation}
 The entanglement entropy of the excited  state $ | \psi \rangle $ satisfies the
 volume law, while that of  the ground state or all the low energy states satisfy the more common area law.

 Noow one can use Eq.\ref{eth} to evaluate the diagonal energy level shift on the chaotic side
 due to the integrable perturbation $ O= H_2 $.
 Taking the complex SYK model at half-filling for an example, assuming the temperatures corresponding to the two
 NN bulk state $ |\psi_1 \rangle, |\psi_2 \rangle $ in the chaotic side are $ \beta_1, \beta_2 $, then
 one can see:
\begin{equation}
  \langle \psi_1 | H_4 | \psi_1 \rangle -  \langle \psi_2 | H_4 | \psi_2 \rangle
  =\frac{1}{Z} Tr H^2_{4} e^{-\beta_1 H_4 } \delta \beta
\label{deltaT}
\end{equation}
   Equating Eq.\ref{deltaT} to Eq.\ref{dualC} leads to $ \delta \beta=\beta_1- \beta_2 \sim  N^{3/2} 2^{-N} $.
   Then one can immediately see  the diagonal energy level shift:
\begin{eqnarray}
  \langle \psi_1 | H_2 | \psi_1 \rangle -  \langle \psi_2 | H_2 | \psi_2 \rangle
  & = & \frac{1}{Z} Tr H_2 H_{4} e^{-\beta_1 H_4 } \delta \beta  \nonumber  \\
  & \sim &  N^{3/2} 2^{-N}
\label{diagC}
\end{eqnarray}
    which is very tiny. Then one may focus on the off-diagonal matrix element  $  \langle \psi_1 | H_2 | \psi_2 \rangle $.
    By using the Cauchy inequality $ |  \langle \psi_1 | O | \psi_2 \rangle |^2
    < \langle \psi_1 | O | \psi_1 \rangle  \langle \psi_2 | O | \psi_2 \rangle  $, one can establish the bound:
\begin{equation}
    |  \langle \psi_1 | H_2 | \psi_2 \rangle | <  Tr H_2 e^{-\beta_1 H_4 }/Z
\label{H12C}
\end{equation}
   which maybe useful to evaluate the dual form Eq.\ref{dualC} in Sec. V.


 Because ETH focus on excited states, so it should be closely related to RMT.
 Indeed,  for interacting many body systems, the quantum chaos imply the ETH or vice versa \cite{ETHrev}.
 The results achieved in this paper show that the ETH of $ q=4 $ SYK should be preserved
 in all the plateau regimes in Fig.3-8 satisfying a WD class in the hybrid SYK  models.
 While the KAM theorem implies the violation of ETH or broken ergodicity.
 The transition regimes between different plateaus may not satisfy ETH either.

 The ETH only applies to the bulk states, so it has a very serious limitation: it has no saying
 on ground state and the low energy excitations which can be defined
 as $ \lim_{V \rightarrow \infty} \frac{E-E_0}{V} = 0 $. They are nothing but the edge states.
 So the conjecture that a single bulk eigenstate encodes the information of the
 full Hamiltonian \cite{ETHsingle} clearly fails on the edge states.
 A complete understanding of the system needs not only
 the knowledge of the bulk states in a statistics way by a RMT, but also the edge states by more
 quantitative approach such as 1/N expansion or RG.


\subsection{ Comments on some early works on type-I hybrid Majorana SYK models }

  In this section, we comment on previous works  \cite{stupid,new} on the type-I hybrid SYK model
  and also point out their main differences than our work.

The special GUE case (a) and (c) in Fig.\ref{MajoranaSYK_Type1_Even}
(but not the other two cases of GOE in (b) and GSE in (d) )
in type-I Majorana fermion hybrid SYK  Eq.\ref{mix1} at even $N$
was studied in  \cite{stupid}. However,
the most interesting case of the interruption of GUE in the intermediate ranges of $ K/J $ in the other two cases (b) and (d) is absent in this special GUE case (a) and (c).
Furthermore, these authors in-correctly interpret the CIT
as a true QPT at $T=0$ or a classical phase transition at $T>0$.
So it does not correspond to the Hawking-Page transition in its bulk gravity dual as claimed in this work.
It was known that the Hawking-Page transition in the bulk
may be dual to a true quark-gluon  confinement to deconfinement transition at $T>0$ in the boundary.
The dramatic differences between the CIT characterized
by RMT  at a finite $ N $ and a true quantum or a classical phase transitions
characterized by RG at $ N=\infty $  was stressed in the last subsection.
The finite size scalings to locate a quantum critical point (QCP) only apply to the later,
not to the former. Unfortunately, the authors in  \cite{stupid} still tried to fit their data to a finite size scaling without success.

After submitting the first version of our work to arXiv,
we got to know the authors in \cite{new}
have also studied the other two cases of (b) and (d).
Furthermore, they also evaluated the Thouless energy scale $ E_{th}=N^2 \Delta $
where $ \Delta $ is the average many body energy level spacing beyond which the RMT breaks down \cite{randomM0}.
However, this reference did not (1) introduce the new NNN ratio $r'$ (2) do the $ odd $ N case
(3) study the type II cases (4) address the possible deep relations between
the RG at $ N =\infty $, $1/N $ expansion at a finite $ N $ and the RMT at a finite $ N $.
As shown in Sec.III-A,Sec.IV-A, the odd number of sites are in different classes in both classifications and ED \cite{unknown}.
In the GSE case in Fig.3-5,
which has the double degeneracy at the $ q=4 $ side,
the new NNN ratio $r'$ must be used to describe the stability regime
of the quantum chaos in the GSE side and also describe the whole crossover from GSE to the GUE,
then the CIT from GUE to the KAM side with the Poisson distribution.
The new NNN ratio must be used to describe the KAM theorem in the integrable side
when there is a double degeneracy in the integrable side as are the cases in Fig.6b, Fig.7b,c and Fig.8a
in the type-II hybrid SYK models.

\section{Conclusions and perspectives }

 RG is a semi-group which may not have an inverse. In the coarse graining process, some information gets lost,
 the self similar phenomena start to emerge only in the low energy limit. What kind of information is lost ?
 We believe what is lost is the RMT information.
 It is worth to point out that in relativistic  quantum field theories, the low energy and high energy levels are closely related
 due to the Lorentz invariance.
 So the RG can be equivalently performed by removing the IR or UV divergencies. This procedure is well established by
 the dimensional regularization in relativistic quantum field theories. It has also been applied to non-relativistic quantum field theories
 to describe the superfluid to Mott transition \cite{ye1} with the long range Coulomb interaction
 and also the quantum Hall plateau-plateau transition in a periodic potential with the long range Coulomb interaction
 \cite{ye2,ye3}. We expect the bulk energy levels of these systems are also described by the RMT. Similarly,
 the QCD is described by the asymptotical freedoms in the high energy ( or short distance ) limit, but its bulk energy levels
 are described by the RMT \cite{QCD-a,QCD-b,QCD-c}.


   Here we introduced a new universal ratio which is the ratio of the next nearest neighbour (NNN) energy level spacing
   to characterize the random matrix behaviours. It must be used when there is a double ( $ d=2 $ ) degeneracy near the chaotic side or
   near the integrable side.   This new universal ratio is particular useful when
   numerically characterizing the KAM regime in the integrable side and  the stability of quantum chaos in the chaotic regime.
   In Sec.V, we only present some very preliminary results on the analytical scaling forms of the KAM and its dual form.
   Their complete and rigorous forms will be given in a separate publication \cite{math}.
   If there are higher order such as $ d=3,4.... $  degeneracy such  near the chaotic side or
   near the integrable side, then one may need to introduce more universal ratios such as the NNNN, NNNNN ratios
   or the whole series whose physical meanings remain to be explored. This is similar to Tensor category, upto higher order tensor category
   are needed to characterize the topological phases \cite{wen}.

   In a recent work\cite{color}, we studied quantum chaos in 2- or 4-colored SYK models and also CIT in
2- or 4-colored hybrid SYK models. These colored SYK models provide concrete examples of classifying quantum chaos
in a system with multiple conserved quantities which show richer RMT classes \cite{colorMBL} than the conventional SYK models.
The NNN ratio should also find its applications in numerically studying KAM and its dual form in the 2- or 4-colored hybrid SYK models.
It may also be interesting to study the ELS of the two indices SYK model \cite{me}
   with the two large numbers: $ N $ (the number of sites) and the $ M $ (the group of $ O(M) $ or $ SU(M)$).
   Depending on the relations between $ N $ and $ M $, it may show either chaotic QSL or symmetry-broken QSG ground state.

As shown in \cite{pure}, the stability of quantum chaos of black holes
against a constant $ q=2 $ SYK terms may be used to explore the interior
behind the black hole horizon.
Implications of the results achieved here on the bulk gravity side
or on quantum error corrections to AdS/CFT need to be explored \cite{QEC}.

Fig.9b1 and Fig.9b2 show that the $ q=4 $ chaotic SYK CFT fixed point and
 the $ (q=2)^2 $ integral CFT flow towards each other with $ \pm $ sign respectively.
 In 2d CFT, Zamoloddchikov's c-theorem states that RG flows only from a CFT with central charge $ c_1 $
 to another  one with $ c_2 < c_1 $. One can construct a c-function which monotonically decrease from $ c_1 $ to $ c_2 $.
 The 2d C-theorem can also be extended to higher dimensions through F-theorem in odd dimension or a-theorem in even dimension \cite{holoRG}.
 In the 2d boundary CFT (BCFT), there is a also a g-theorem \cite{kondo1,kondo1g,kondo2,kondo3,kondo4} which states that there is only one way RG flow from
 a BCFT with the boundary degeneracy $ g_1 $ to another  one with $ g_2 < g_1 $.
 However, there is no such a c-theorem in a 1d CFT.
 This maybe special to the 1d CFT where one can only associate a Lyapunov exponent $ \lambda_L $ when away from the CFT due to
 a leading irrelevant operator ( so called NCFT$_1$ )
 instead of a central charge $ c $ to a CFT fixed point. This may also be related to the distinction of $ NAdS_2/NCFT_1 $
 from its higher dimension counterparts. It remains important to explore further the possible deep mechanism on the two ways RG flow
 between the FL and NFL fixed points in 1d CFT.



\acknowledgments
J. Ye thank C. Xu for helpful discussions.
F.S and J.Y thank Prof. Gang Tian and Prof. Congjun Wu  for the hospitality during their visit at the West Lake university;
also thank Prof. Wei Ku for the hospitality during their visit at the Tsung Dao Lee Institute in Shanghai, China.
F.S and J.Y acknowledge AFOSR FA9550-16-1-0412 for supports. Y.Y thank Y.-Z. You for sharing the ED code used in Ref.\cite{MBLSPT}.
This research at KITP was supported in part by the National Science Foundation
under Grant No. NSF PHY-1748958.
WL was supported by the NKBRSFC under grants Nos. 2011CB921502, 2012CB821305, NSFC under grants Nos. 61227902, 61378017, 11311120053.

\appendix

\section{The hybrid of $ q=2 $ and $ q=4 $  bosonic  SYK }

    This appendix was cited in Sec.VI-A.  The procedures for fermions presented in the main text can also be applied to study the $ q=2 $ and $ q=4 $
    hybrid bosonic SYK:
\begin{equation}
H_{B}=  \sum^{N}_{i<j,k<l} J_{ij;kl} b^{\dagger}_{i} b^{\dagger}_{j} b_k b_l +  \sum_{i<j}
K_{ij} b^{\dagger}_{i} b_{j}- \mu q_b
\label{mix3}
\end{equation}
  where, in general, $ J_{ij;kl}=J_{ji;kl}, J_{ij;kl}=J_{ij;lk}, J^{*}_{ij;kl}= J_{kl;ij} $ and
  $ \langle |J_{ij;kl}|^2 \rangle= 3! J^2/N^3 $. $ q_b= \sum_{i} ( b^{\dagger}_i b_i -1/2) $ is just
  the boson analog of Eq.\ref{qc}.
  Following \cite{CSYKnum}, {\sl we take the four site indices $ i,j; k,l$ are all different \cite{matter} to keep
  the PH symmetry explicit at $ \mu=0 $. }.
  $ K^{*}_{ij} =K_{ji} $ is  a Hermitian matrix satisfying  $ \langle K_{ij} \rangle=0,  \langle |K_{ij}|^2 \rangle=  K^2/N $.

  In the $ K/J=0 $ limit,  the $ q=4 $  bosonic  SYK was studied by the ED in \cite{CSYKnum},
  a Quantum spin glass (QSG) ground state was expected in the thermodynamic limit $ N=\infty $.
  One can define the particle-hole symmetry operator to be $ P= K \prod^{N}_{i=1} ( b^{\dagger}_i + b_i ) $. The boson charge
  operator is $ Q_b= \sum_{i} b^{\dagger}_i b_i  $.
   It is easy to show $ P^2=1,  P b_i P= b^{\dagger}_i, P b^{\dagger}_i P= b_i,
   P Q_b P= N- Q_b $. $ [P, H_{M4}]=0 $.
   For $ N $ even, at half filling $ q_b=0 $, it is in GOE.
   However, as long as $ q_b \neq 0 $,  there is no PH symmetry anymore,
   it is in GUE regardless of the value of $ N  $ is even or odd.

  Now we apply the PH transformation to the  bosonic hybrid SYK model Eq.\ref{mix3}.
  In contrast to the fermionic hybrid SYK models, $ [P,H_2]=0 $, so the PH symmetry is preserved in the  hybrid bosonic SYK model.
  In the $ J/K=0 $ limit, we expect that the ELS for $ q=2 $ bosonic SYK is the same as $ q=4 $ bosonic SYK:
  when $ q_b =0 $, it is in GOE, when  $ q_b \neq 0 $, it is in GUE.
  This is in sharp contrast to the $ q=2 $ fermionic SYK which is non-interacting, so integrable.
  While  the $ q=2 $ bosonic SYK is interacting
  (the bosons on the same site behaves as fermions, but different sites as bosons),
  so non-integrable and is already a quantum chaotic system.

\begin{figure}
\centering
\includegraphics[width=\linewidth]{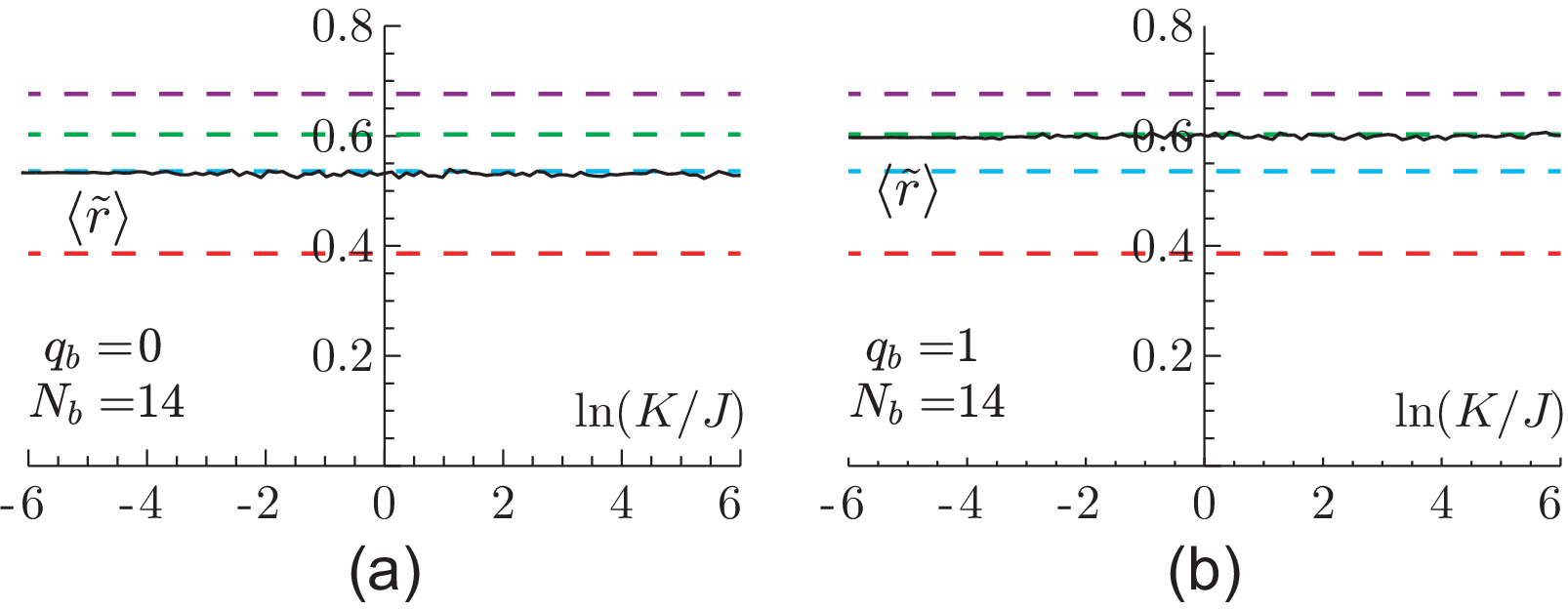}
\caption{ELS of the hybrid  bosonic SYK.
(a) For $ N_b $ even, at the half filling $ q_b=0 $, it is always in GOE for any ratio of $ K/J $.
(b) Away from the half filling $ q_b \neq 0 $, it is always in GUE for any ratio of $ K/J $, regardless of $ N_b $ is even or odd.
    In contrast to the fermion cases presented in the main text, there is no CIT in the bosonic hybrid SYK.
	However, it is not known if there is a QPT between the bosonic QSG$_2$ and QSG$_4$ in the ground state. }
\label{bcross}
\end{figure}


   So we expect the ELS of the hybrid bosonic SYK stays the same from $ q=4 $, all the way down to $ q=2 $.
   This is indeed confirmed by our ED results \cite{JWb}
   shown in Fig.\ref{bcross}, there are only two cases here
   (a) For $ N_b $ even, at the half filling $ q_b=0 $, it is always in GOE for any ratio of $ K/J $.
   (b) Away from the half filling $ q_b \neq 0 $, it is always in GUE for any ratio of $ K/J $, regardless of $ N_b $ is even or odd.

  In short, in contrast to all the hybrid fermionic models discussed in the main text, the KAM theorem does not apply in
  the bosonic hybrid model where  there is no CIT. Here we only focused on the ELS of bulk spectrum.
  It supports the claim made in Sec.VI-A on the relation between the
  edge versus bulk energy levels and belong to the class A in the BSCFS.

\section{ The many-body density of states of $ q=2 $ Majorana and $ q=2 $ complex  SYK models}

In this appendix, we provide the many body energy distributions of  the $ q=2 $ Majorana and $ q=2 $ complex  SYK models
which are needed to derive the scaling forms of the KAM theorems for the corresponding hybrid SYK models in Sec. V.
Surprisingly, there is no previous works to discuss the many body energy level distributions of
the $ q=2 $ Majorana or complex SYK models.

\begin{figure}[!htb]
\centering
\includegraphics[width=0.9\linewidth]{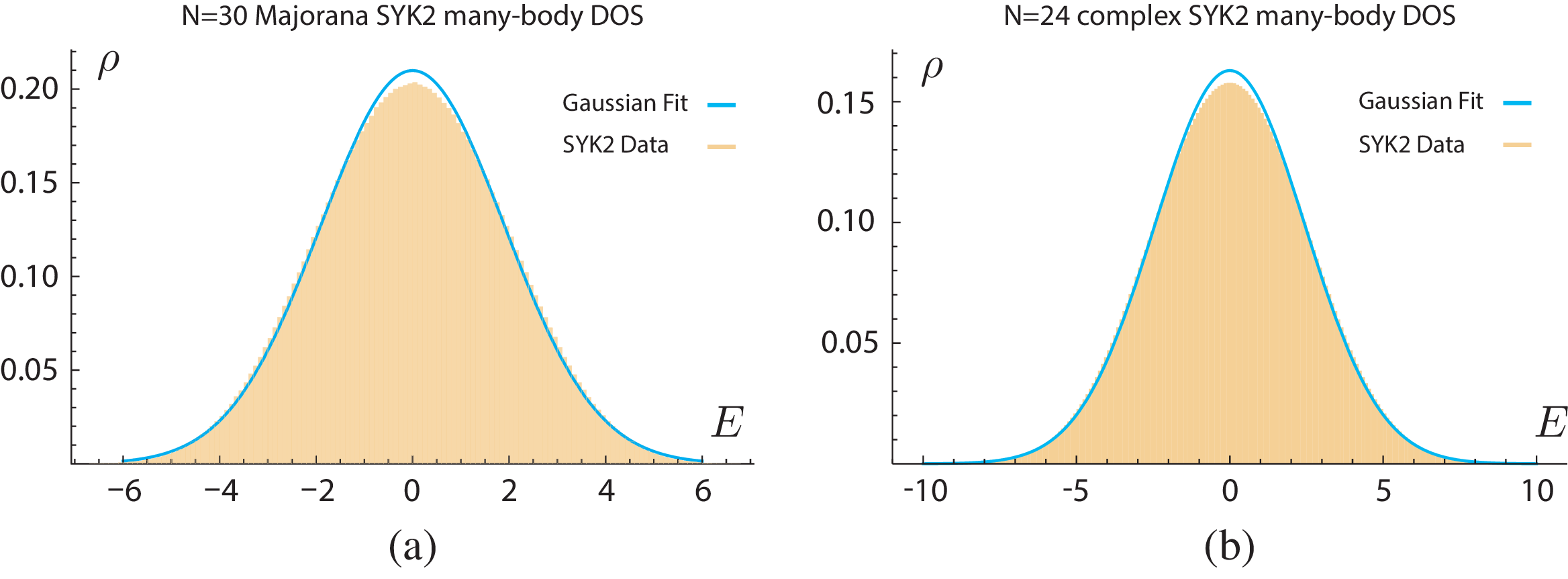}
\caption{ The Many-body density of states of $q=2$ Majorana and complex SYK models: (a) Majorana fermion case at a given parity
(b) complex fermion case at half-filling.
	Both many body DOS are close to be a Gaussian distribution with a small deviation in the center \cite{exactproof}.
However, it is difficult to resolve the precise nature of the band edge numerically. }
\label{type1MB-DOS}
\end{figure}

{\sl 1. $ q=2 $  Majorana SYK model.}

The $q=2$  Majorana SYK model in Eq.11 is defined as $H_\chi= i \sum_{1\leq i<j\leq N} K_{ij}\chi_i\chi_j$,
where $K_{ij}$ is real random number and drawn from the Gaussian distribution with zero mean and $K^2/N$ variance.
So its single particle energy levels fit rigorously the RMT description \cite{Gross} with the matrix size $ L=N $.

One can calculate its all many-body energy levels $\{E_\chi\}$ at a given parity sector
by diagonalizing a $2^{N/2-1}\times 2^{N/2-1}$ sparse matrix (assuming $N$ is even).
Obviously, $\langle E_\chi\rangle=0$.
The data  $\langle E_\chi^2\rangle$ is listed in Table \ref{tab:SKY2-Majorana}.
We find that the many-body energy level statistics (ELS) is a Poisson
and the many-body energy density of states (DOS) $\rho(E_\chi)$ satisfies a Gaussian distribution with zero mean
and variance $\sigma^2=0.127N-0.157\approx N/8$. This value matches the analytic estimate
of the second moment $ TrH^k/Tr1=[ \frac{\langle J^2 \rangle}{2^q} C^q_N]^{k/2} $ \cite{randomM}.
Putting $ k=2, q=2 $ and $ \langle K^2 \rangle=K^2/N $, we find it is $ K^2 N/8 $.
The fitted Gaussian distribution and $\rho(E_\chi)$ are shown in Fig.\ref{type1MB-DOS}a.

\renewcommand{\arraystretch}{1.3}
\renewcommand\tabcolsep{8pt}
\begin{table}[!htb]
\caption{The $\langle E_\chi^2\rangle$ of the $ q=2 $ Majorana SYK
in the even parity sector averaged over $1000$ samples.
We choose $K=1$ to perform the ED.
The data of $\langle E_\chi^2\rangle$ show a linear dependence on $N$,
and give $\langle E_\chi^2\rangle=0.127N-0.157$. }
\begin{tabular}{ c|*{11}{c} }
\hline
$N_\chi$				&12	&14	&16	&18	&20	\\
\hline
$\langle E_\chi^2\rangle$	&1.354	&1.631	&1.864	&2.129	&2.397	\\
\hline
\hline
$N_\chi$			   &22	&24	&26	&28	&30\\
\hline
$\langle E_\chi^2\rangle$	&2.624	&2.887	&3.085	&3.391	&3.665\\
\hline
\end{tabular}
\label{tab:SKY2-Majorana}
\end{table}

{\sl 2. $ q=2 $  Complex SYK model.}

The  $q=2$  complex SYK model in Eq.14 is defined as $H_c=\sum_{1\leq i < j\leq N} K_{ij}c_i^\dagger c_j$,
where $K_{ij}=K_{ji}^*$ is  random complex number drawn from the Gaussian distribution
with $\langle K_{ij}\rangle=0$ and $\langle |K_{ij}|^2\rangle=K^2/N$.
The fermion number $ Q_c=\sum_i c_i^\dagger c_i=q $ is conserved.
So its single particle energy levels fit rigorously the RMT description \cite{Gross} with the matrix size $ L=N $.

One can calculate its all many-body energy levels $\{E_c\}$ in the half-filling sector ( $ q=N/2 $ for $ N $ even )
by diagonalizing a $ C^{N/2}_N \times  C^{N/2}_N $ sparse matrix (assuming $N$ is even).
Obviously, $\langle E_\chi\rangle=0$.
The data of $\langle E_\chi^2\rangle$ is listed in Table \ref{tab:SKY2-complex}.
We find the many-body ELS is a Poisson
and many-body energy DOS $\rho(E_c)$ satisfies a Gaussian distribution with zero mean
and $\sigma^2=0.250N+0.002\approx N/4$ variance.
The fitted Gaussian distribution and  $\rho(E_c)$ are shown in Fig.\ref{type1MB-DOS}b.

\renewcommand{\arraystretch}{1.3}
\renewcommand\tabcolsep{8pt}
\begin{table}[!htb]
\caption{The  $\langle E_c^2\rangle$ of the $ q=2 $ complex SYK at half filling averaged over $10000$ samples.
We choose $K=1$ to perform the ED.
The data of $\langle E_c^2\rangle$ show a linear dependence on $N$,
and give $\langle E_c^2\rangle=0.250N+0.002$.}
\begin{tabular}{ c|*{8}{c} }
\hline
$N_c$			&10	&12	&14	&16	\\
\hline
$\langle E_c^2\rangle$	&2.499	&3.001	&3.501	&4.001	\\
\hline
\hline
$N_c$			&18	&20	&22	&24\\
\hline
$\langle E_c^2\rangle$   &4.501	&5.004	&5.503	&5.998\\
\hline
\end{tabular}
\label{tab:SKY2-complex}
\end{table}

\section{ The many-body density of states of $ q=4 $ complex  SYK models}

\begin{figure}[b]
\centering
\includegraphics[width=0.5\linewidth]{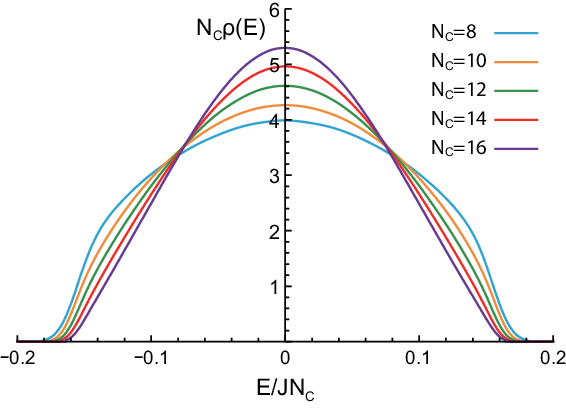}
\caption{The normalized many body DOS $\rho(E)$ for the half-filling sector of complex SYK model
with $N_c=8,10,12,14,16$. The numbers of samples are
50000 ($N_c=8$), 10000 ($N_c=10$), 2000 ($N_c=12$), 400 ($N_c=14$), 10 ($N_c=16$).  }
\label{complexdos}
\end{figure}

 The many body DOS for the $ q=4 $ Majorana SYK was studied by $ 1/N $ in \cite{Mald} and ED in \cite{randomM}.
 It was found that globally it is like Gaussian with the variance $ \sqrt{ J^2N/64} $, locally look like the semi-circle DOS with $ \sqrt{E-E_0} $
 behaviour near the band edge ( either low or high ).
 The many body DOS for the $ q=4 $ complex SYK is shown in Fig.\ref{complexdos}.

   At fixed $ q=4 $, as $ N $ gets large, the many body DOS of Majorana or complex SYK
   globally approaches the Gaussian with a width $ \sigma \sim \sqrt{N} $. However, near the band edge $ E_0 \sim N $,
   it shows the $ \sqrt{E-E_0} $ behaviour. So near the band edge,
   it locally behaves as a semi-circle. The $ 1/N $ expansion at a fixed $ q $ at the temperatures  $ 1< \beta J < N $ can only resolve the local $ \sqrt{E-E_0} $ behaviour near the band edge, but not the global behaviour.
   In the double scaling limit $ N \rightarrow \infty $ and $ q \rightarrow \infty $, but keep $ \lambda= q^2/N $ fixed,
   the action can be mapped to a 2d Liouville CFT in the kinetic space with the central charge $ c \sim N/q^2=1/\lambda $ which
   is solvable at all energy scales \cite{double1}.
   The DOS was shown to be a Gaussian when $ c \gg 1 $, a semi-circle  when $ c \ll 1 $,
   a more complicated form when $ c \sim 1 $  ( see also the footnote [46] in Ref.\cite{period}   ).
   However,  only in the triple scaling limit $ \lambda \rightarrow 0 $ ( namely, the central charge  $ c \rightarrow \infty $ )
   and the low energy limit $ E-E_0 \rightarrow 0 $, at a fixed $ (E-E_0)/J \lambda = z $, it reduces to the 1d Schwarzian
   which is the low energy limit of the SYK model.

\section{ The edge ( low and high energy) states versus the bulk ( intermediate energy ) states in $ q=2 $ and $ q=4 $
complex SYK }

In this appendix, we aim to demonstrate the connections and differences between the edge and bulk states in the Fock space.
They can be studied by studied by $ 1/N $ expansion or RG and RMT respectively. This appendix is heavily cited in Sec.VI.

{\sl 1. $ q=2 $  Complex SYK model.}

\begin{figure}[t]
\centering
\includegraphics[width=0.9\linewidth]{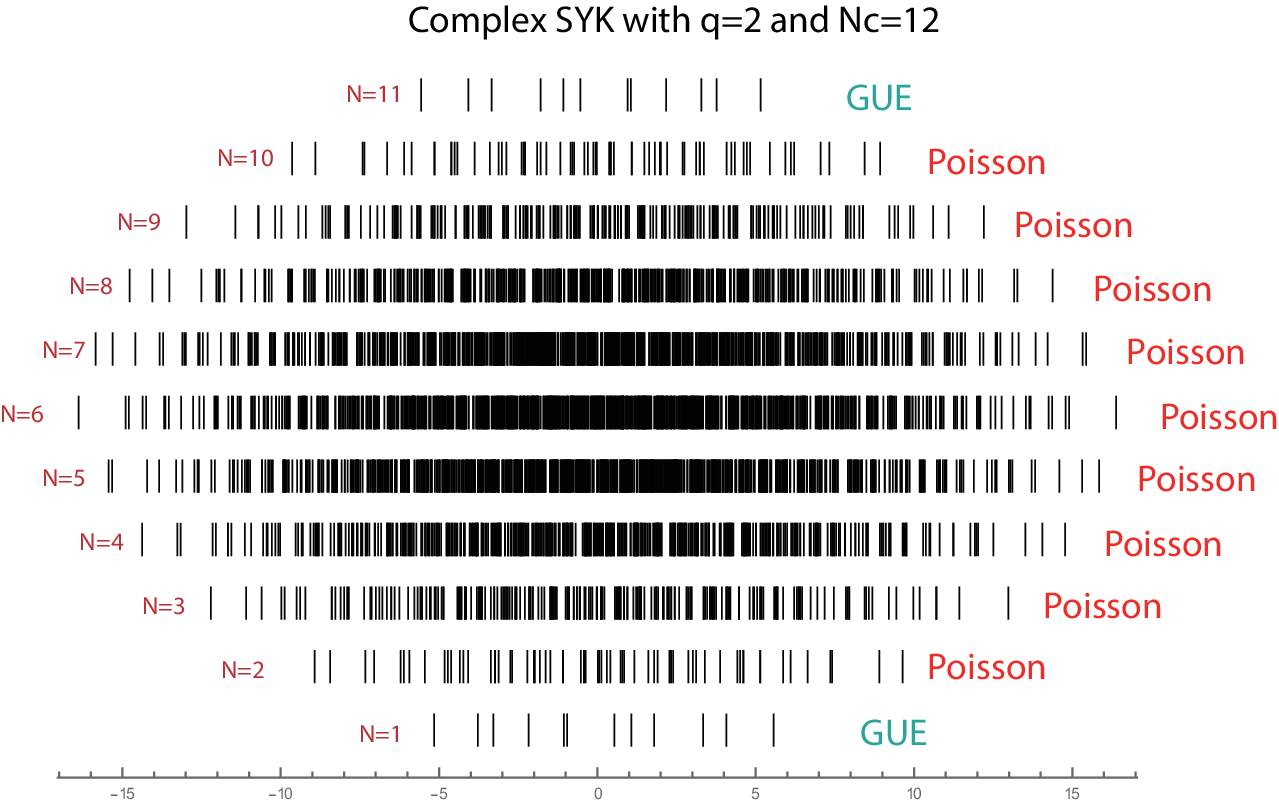}
\caption{ The edge and bulk energy levels of  $q=2$ complex SYK at $N_c=12$.
The energy levels are labeled by conserved quantity ( the number of fermion $ Q_c $ ).
Only the $Q_c=1,11$ sectors satisfy GUE, the others are Poisson.
Note: $Q_c=0$ and $Q_c =12$ are two trivial sectors. The many body DOS at the half-filling  is given in Fig.\ref{type1MB-DOS}b.}
\label{q2complex}
\end{figure}

   The $ q=2 $ complex SYK was written in Eq.\ref{mix2}:
\begin{equation}
   H_2= \sum^N_{i<j} K_{ij} c^{\dagger}_{i} c_{j}- \mu q_c
\label{hc2}
\end{equation}
   Instead of using the grand-canonical ensemble, we choose canonical one with a fixed fermion number
   $ Q_c= \sum^{N}_{i=1} c^{\dagger}_i c_i $. Then, the total Hilbert space can be decomposed into $ N $ blocks as $ 2^N= \sum^N_{q=0} C^q_N $.

For $ Q_c=1 $, it is nothing but the single-particle sector with the dimension $ C^{1}_{N}=N $.
Under the PH transformation $ P $, it is mapped to its particle-hole conjugate sector $ C^{N-1}_{N}=N $
which is nothing but the single-hole sector. Surprisingly and interestingly, only the single-particle and the single-hole
sector  are chaotic in the GUE, all the other multi-particle sectors $ q \geq 2 $ are integrable in the Possionian ( Fig.\ref{q2complex} ).

   As shown in Sec.III, $ \{P,H_2 \}=0 $, $ P $ also maps $ Q_c $ to $ N_c-Q_c $.
   For $ N_c $ even, at the half-filling $ Q_c=N_c/2 $,
   so $ H_2 $ has a mirror symmetry for any realization of $  K_{ij} $ with respect to $ E=0 $. This is indeed the case
   for $ N_c=12,Q_c=6 $ in Fig.\ref{q2complex}. However, the exact mirror symmetry is absent away from the half-filling.
   But as argued below Eq.\ref{h4}, there is still an approximate mirror symmetry even away from the half-filling.
   Let us focus at the half-filling sector in the following.

   As shown in Sec.V, the low energy spacing $ \sim 1/N $ which stands for the low energy quasi-particle excitations, due to the exact mirror symmetry, the high energy spacing
   is also $\sim 1/N $  which stands for the high energy quasi-particle excitations.
   The bulk energy spacing is much smaller $ \sim 1/C^{N/2}_{N} \sim \sqrt{N} 2^{-N} $ at a large $ N $.
   The KAM theorem is
   determined by the bulk energy spacing. All the bulk energy levels are described by the RMT in the Possion statistics.
   The bulk energy level statistics is quite in-sensitive to the edge energy levels. Namely,
   incorporating or throwing away these edge ( low or high energy ) levels will not affect the bulk Possion statistics.

{\sl 2. $ q=4 $  Complex SYK model.}

\begin{figure}[b]
\centering
\includegraphics[width=0.9\linewidth]{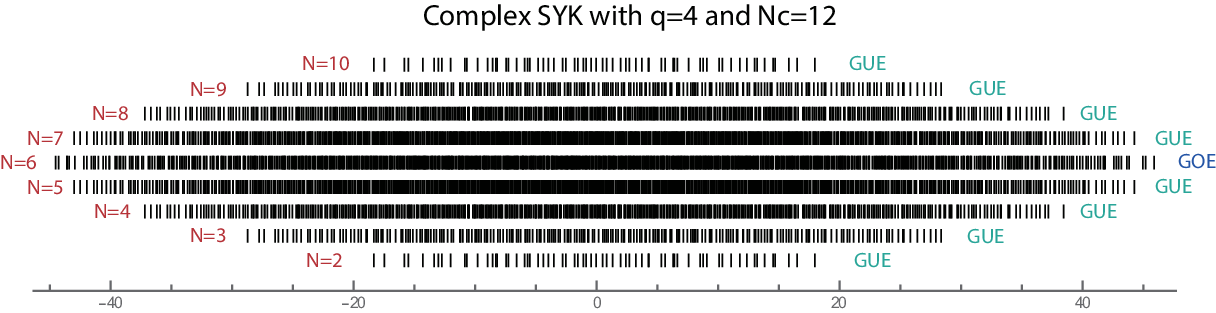}
\caption{The edge and bulk energy levels of $q=4$ complex SYK and $N_c=12$.
The energy levels are labeled by conserved quantity, number of fermion $ Q_c $.
Only the $ Q_c=6$ sectors satisfy GOE, others are GUE.
Note: For $ Q_c=1, N_c-1 $ stand  for a single particle or hole, it is an exact zero modes with the degeneracy $ N_c $.
$Q_c=0, 12$ are two trivial sectors. The many body DOS at the half-filling  is given in Fig.\ref{complexdos}. }
\label{q4complex}
\end{figure}

 The $ q=4 $ complex SYK was also written in Eq.\ref{mix2}:
\begin{equation}
H_{4}=  \sum^{N}_{i<j,k<l} J_{ij;kl} c^{\dagger}_{i} c^{\dagger}_{j} c_k c_l - \mu q_c
\label{hc4}
\end{equation}

As shown in Sec.III, $ [P,H_4 ]=0 $, so in contrast to $ H_2 $, $ H_4 $ does not
have a mirror symmetry for any given random realization of $  J_{ij;kl} $.
However, it still have an approximate mirror symmetry for any given realization of $ J $ due to the following argument
similar to Eq.\ref{pmeq}:
\begin{equation}
 - H_{4}= - \sum^{N}_{i<j,k<l} J_{ij;kl} c^{\dagger}_{i} c^{\dagger}_{j} c_k c_l =
     \sum^{N}_{i<j,k<l} J^{\prime}_{ij;kl} c^{\dagger}_{i} c^{\dagger}_{j} c_k c_l
\label{h4}
\end{equation}
 where $ J^{\prime}_{ij;kl}=-J_{ij;kl} $.
 Note that we are still confining to the $ Q_c= \sum^{N}_{i=1} c^{\dagger}_i c_i $ sector.
 Obviously, $ J^{\prime} $ and $ J $ satisfy the same distribution.
 At a large enough $ N_c$, it is self-averaging, so there is still an approximate mirror symmetry at any filling
 for a given random realization of $  J_{ij;kl} $.

This is indeed the case shown in Fig.\ref{q4complex}.
  As shown in Sec.VI, the low energy spacing $ \sim e^{-s_0N }$, due to the approximate mirror symmetry, the high energy spacing
   is also $\sim e^{-s_0N } $. The bulk energy spacing is much smaller $ \sim 1/C^{N/2}_{N} \sim \sqrt{N} 2^{-N} $. The KAM theorem is
   determined by the bulk energy spacing. All the bulk energy levels are described by the RMT in the GOE.
   The bulk energy level statistics is quite in-sensitive to the edge energy levels. Namely,
   incorporating or throwing away these edge ( low or high energy ) levels will not affect the bulk GOE statistics.


 It remains interesting to examine how the edge and bulk states evolve from the $q=2 $ side in Fig.\ref{q2complex}
 to that in the $ q=4 $ side in Fig.\ref{q4complex} in the Type-I and Type-II hybrid complex SYK models.

\bibliographystyle{apsrev4-1}

\begin{thebibliography}{99}
\bibitem{TR}
In fact, equivalently, one can also define the PH symmetry operator to be
$R=P(-1)^{Q_c}=(-1)^{N_c}(-1)^{Q_c}P=K\prod^{N_c}_{i=1}(c^{\dagger}_i-c_i)$.
It is easy to show $R^2= (-1)^{[\frac{N_c}{2}]+N_c}$,
$R c_i R= (-1)^{N_c-1 } \eta c^{\dagger}_i$,
$R c^{\dagger}_i R= (-1)^{N_c-1 } \eta c_i$,
$R \chi_i R= (-1)^{N_c-1 } \eta \chi_i $
where $ \eta=(-1)^{[\frac{N_c-1}{2}]} $.
The total number of fermions $ Q_c= \sum_i c^{\dagger}_i  c_i $.
It is not a conserved quantity, but the parity $ (-1)^{Q_c} $ is in $H_{M4}$.
Then  $ R Q_c R^{-1}= N_c-Q_c $ which justifies $ R $ as an anti-unitary PH transformation.
Here $ R $ does not provide any new information than $ P $, so is redundant.
But they become important and in-dependent anti-unitary operators
in the $ {\cal N}=1 $ SUSY SYK \cite{randomsusy1,randomsusy2,period}.
For a unified and compact classification scheme on both SYK and $ {\cal N}=1 $ SUSY SYK, see \cite{period}.
\bibitem{evenodd}
Here we choose even site $ \chi_{2i} $ to be real, odd site $ \chi_{2i-1} $ to be pure imaginary.
Then as shown in our recent work \cite{period}, by choosing this way,
when $ N $ is odd, one must use the extended scheme
(namely adding a Majorana fermion at $ \infty $ to enlarge
the Hilbert space twice, also introduce one more conserved quantity: the parity).
However, if one choose the other way around, namely, even site $ \chi_{2i} $ to be pure imaginary, odd site $ \chi_{2i-1} $ to be real,
then  when $ N $ is odd,  one can use either the minimum scheme without adding  a Majorana fermion at $ \infty $ or the extended scheme.
\bibitem{unknown}
It seems that so far, Ref.\cite{MBLSPT} and this work are the only two papers which have discussed the classifications
and presented EDs in the odd $ N $ case. But the two papers still use different anti-unitary operators to do the RMT classifications.
For example, Ref.\cite{MBLSPT} constructed an anti-unitary Time-revresal operator in the many body Hilbert space,
while we did not have such a kind of operator.
It is still not known to the authors what are the relations between the two different sets of anti-unitary opartors.
For a unified minimum classification scheme, see Ref.\cite{period}.
\bibitem{can}
In fact, here we discuss a canonical ensemble instead of a grand canonical ensemble.
So it has a fixed $ N $ instead of a fixed $ \mu $.
\bibitem{noise}
In all the Figs.4-8, we get more noises on the $ q=2 $ side than those on the $ q=4 $ sides.
This is because at a given $ N $, there are only $ N^2 $ random numbers $ K_{ij} $ on the $ q=2 $ side, while there are much more $ N^4 $ random numbers $ J_{ijkl} $ on the $ q=4 $ side.
Of course, all $q$-SYK are quite sparse in RMT sense.
The $ q=2 $ side is just more sparse than the $ q=4 $ side.
\bibitem{JW}
To do the ED, one \cite{MBLSPT,CSYKnum} puts the $q$ complex fermions on a chain
with $ N $ sites $ i=1,2,\cdots,N $, then use the Jordan-Wigner transformation
$ c_i= \sigma^{-}_{i} \prod_{j<i} \sigma^{z}_{j}$,
$c_i= \sigma^{+}_{i} \prod_{j<i} \sigma^{z}_{j}$
(or equivalently $ \chi_{2i}= \sigma^{x}_{i} \prod_{j<i} \sigma^{z}_{j}$,
$\chi_{2i-1}= \sigma^{y}_{i} \prod_{j<i} \sigma^{z}_{j}$)
to map the spinless complex (or Majorana) fermion model to a quantum spin model,
then do the ED in the equivalent quantum spin model.
In this work, we did our ED in 3 independent ways
(1) in the above spin basis after making the Jordan-Wigner transformation.
(2) in the Hilbert space of complex fermions. As discussed in Sec.III-A,
for odd $ N $, we add a Majorana fermion at $ N+1=\infty $.
(3) using the Clifford algebra for both $ N=2k $ even and $ N=2k+1 $ odd case.
We achieved the same ED results in all the three ways.
\bibitem{JWb}
To do the ED, following Ref.\cite{CSYKnum}, one puts the $ Q_b $ hard core bosons on a chain
with $ N $ sites $ i=1,2,\cdots,N $.
The hard core boson case is simpler than its fermionic counterpart,
because one do not need to attach a Jordan-Wigner string of $\sigma_z$, namely,
$b_i=\sigma^{-}_{i}$, $b^{\dagger}_i=\sigma^{+}_{i}$
to map the hard core boson model to a quantum spin model.
\bibitem{AFM}
It was known that the quantum ferromagnetic and anti-ferromagnetic Heisenberg model differs by only a sign. Despite they have dramatically different ground states and low energy excitations,
using the same argument here, one can see they share the same ELS.
\bibitem{matter}
This choice is important when doing ED at a finite size $ N $. However,
choosing the four site indices all different or not makes no difference
in the $ 1/N $ expansion.
\bibitem{finitegap}
In the $ U(1) $ Dicke model \cite{gold}, at any finite $ N $,
the gapless Goldstone mode at $ N=\infty $ was lifted to
a pseudo-Goldstone one with a finite gap $ \Delta \sim D/N $
where $ D $ is the diffusion constant subject to an oscialltion due to the Berry phase.
Then when adding a small CRW term $ \beta= g^{\prime}/g < \beta_{U(1)} \sim 1/N^2 $
smaller than this gap opened by the finite size effect,
then the pseudo-Goldstone mode remains robust.
This is more similar to the $ q=2 $ integrable side
where the quasi-particle exists and has a gap $ \sim 1/N $,
but dramatically different than the $ q=4 $ quantum chaostic side
where the quasi-particle breaks down and has a much smaller gap $ \sim e^{- s_0 N} $
and all the low energy levels are also stronglly repelled.
\bibitem{edgediffer}
The edge states here mean the many body  energy levels near the ground state,
so they are ground state plus low energy excitations.
The edge exponents in the RMT for the 7 classes with a mirror symmetry \cite{period}
mean the asympototic behvaiour of the lowest eigenvalue $ \lambda_1 $ near the origin $ E=0 $
which, in the present case, is one of the bulk states.
In some literatures, the former is called soft edge, the latter is  hard edge.
\bibitem{exactproof}
There should be an exact proof on this Gaussian distribution in the many body DOS,
see \cite{math}.
\end{thebibliography}

\end{document}